\documentclass[sigconf, screen]{acmart}

\AtBeginDocument{%
  }

\setcopyright{acmcopyright}
\copyrightyear{2018}
\acmYear{2018}
\acmDOI{10.1145/3623278.3624764}

\acmConference[ASPLOS '24]{Make sure to enter the correct
  conference title from your rights confirmation emai}{April 27 - May 1,
  2024}{San Diego, CA}
\acmPrice{15.00}
\acmISBN{978-1-4503-XXXX-X/18/06}

\newcommand{\rev}{\textcolor{black}{ }\textcolor{black}}

 \usepackage{braket}
\usepackage{tikz}
\newcommand*\circled[1]{\tikz[baseline=(char.base)]{
            \node[shape=circle,draw,inner sep=0.5pt] (char) {#1};}}

\newcommand{\ignore}[1]{}

\copyrightyear{2023} 
\acmYear{2023} 
\setcopyright{acmlicensed}\acmConference[ASPLOS '23]{28th ACM International Conference on Architectural Support for Programming Languages and Operating Systems, Volume 4}{March 25--29, 2023}{Vancouver, BC, Canada}
\acmBooktitle{28th ACM International Conference on Architectural Support for Programming Languages and Operating Systems, Volume 4 (ASPLOS '23), March 25--29, 2023, Vancouver, BC, Canada}
\acmPrice{15.00}
\acmDOI{10.1145/3623278.3624764}
\acmISBN{979-8-4007-0394-2/23/03}

\begin{document}

\title{VarSaw: Application-tailored Measurement Error
Mitigation for Variational Quantum Algorithms}

\author{Siddharth Dangwal*}
\affiliation{%
  \authornote{Both authors contributed equally.
Correspondence: siddharthdangwal@uchicago.edu, gsravi@umich.edu}
  \institution{University of Chicago}
  \country{USA}
}

\author{Gokul Subramanian Ravi*}
\affiliation{%
  \institution{University of Michigan}
  \country{USA}
}

\author{Poulami Das}
\affiliation{%
  \institution{The University of Texas at Austin}
  \country{USA}
}

\author{Kaitlin N. Smith}
\affiliation{%
  \institution{Infleqtion}
  \country{USA}
}

\author{Jonathan M. Baker}
\affiliation{%
  \institution{The University of Texas at Austin}
  \country{USA}
}

\author{Frederic T. Chong}
\affiliation{%
  \institution{University of Chicago}
  \country{USA}
}

\renewcommand{\shortauthors}{Dangwal et al.}


\begin{abstract}

For potential quantum advantage, Variational Quantum Algorithms (VQAs) need high accuracy beyond the capability of today's NISQ devices, and thus will benefit from error mitigation. 
In this work we are interested in mitigating measurement errors which occur during qubit measurements after circuit execution and tend to be the most error-prone operations, especially detrimental to VQAs.
Prior work, JigSaw, has shown that measuring only small subsets of circuit qubits at a time and collecting results across all such `subset' circuits can reduce measurement errors.
Then, running the entire (`global') original circuit and {extracting the qubit-qubit measurement} correlations can be used in conjunction with the subsets to construct a high-fidelity output distribution of the original circuit. 
Unfortunately, the execution cost of JigSaw scales polynomially in the number of qubits in the circuit, and when compounded by the number of circuits and iterations in VQAs, the resulting execution cost quickly turns insurmountable.  

To combat this, we propose VarSaw, which improves JigSaw in an application-tailored manner, by identifying considerable redundancy in the JigSaw approach for VQAs: spatial redundancy across subsets from different VQA circuits and temporal redundancy across globals from different VQA iterations.
VarSaw then eliminates these forms of redundancy by \emph{commuting} the subset circuits and \emph{selectively executing} the global circuits, reducing computational cost ({in terms of the number of circuits executed}) over naive JigSaw for VQA by 25x on average and up to 1000x, for the same VQA accuracy.
Further, it can recover, on average, 45\% of the infidelity from measurement errors in the noisy VQA baseline. 
Finally, it improves fidelity by 55\%, on average, over JigSaw for a fixed computational budget. VarSaw can be accessed here: \url{https://github.com/siddharthdangwal/VarSaw}

\end{abstract}




\received{22 October 2022}
\received[revised]{2 March 2023}
\received[accepted]{27 April 2023}


\maketitle

\section{Introduction}

Quantum computers leverage superposition, interference, and entanglement to give them potentially significant computing advantage in chemistry~\cite{kandala2017hardware}, optimization~\cite{moll2018quantum}, machine learning~\cite{biamonte2017quantum} and other domains of critical interest.

In near-term quantum computing, called Noisy Intermediate-Scale Quantum (NISQ), we expect to work  with  {machines that comprise} 100-1000s of imperfect qubits ~\cite{preskill2018quantum}.

NISQ devices suffer from high error rates in the form of state preparation and measurement (SPAM) errors, gate errors, qubit decoherence, crosstalk, etc.
One of the most promising quantum tasks for near-term quantum advantage in the NISQ era are variational quantum algorithms (VQAs).
They have wide application in approximation~\cite{moll2018quantum}, chemistry~\cite{peruzzo2014variational} {etc, and are usually} designed as minimization problems.
VQAs are hybrid quantum-classical algorithms iteratively run a parameterized quantum circuit (called an ansatz) on the quantum machine.
The QC parameters are optimized each iteration by a classical tuner/optimizer to try and approach the global minimum of the variational objective function.
The classical optimizer attempts to adjust the QC to the noise characteristics of the quantum device, and hence, \emph{in theory}, gives VQAs the potential for quantum advantage even on noisy machines.
But \emph{in practice}, noise has prevented current quantum computers from surpassing the capabilities of classical computers in almost all applications, including VQAs.
Considering the potential for VQAs on NISQ machines~\cite{peruzzo2014variational,farhi2014quantum}, it is imperative to explore techniques to maximize their quality of execution on today's machines.

\begin{figure}[t]
\centering
\includegraphics[width=0.9\columnwidth,trim={0cm 0cm 0cm 0cm},clip]{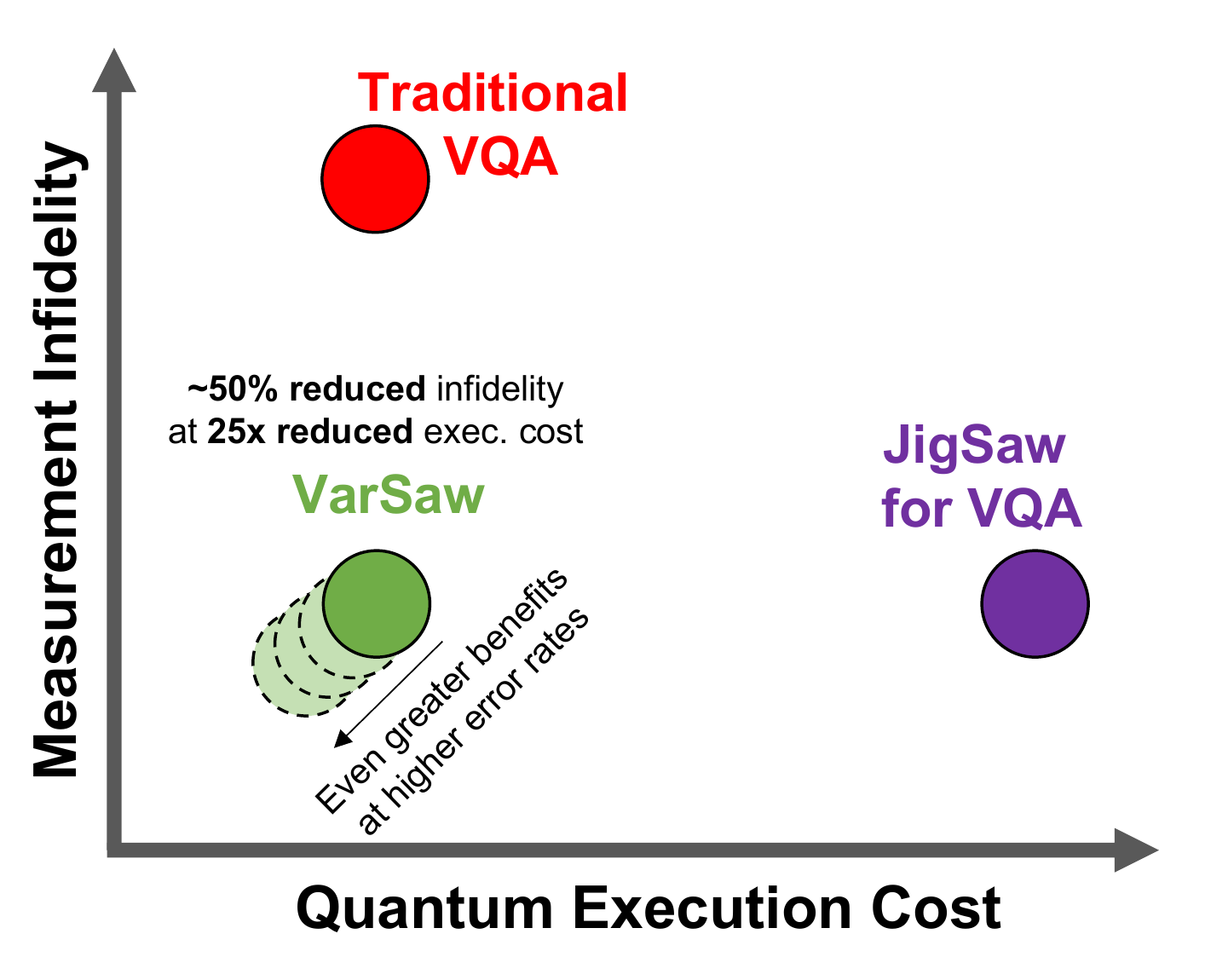}
\caption{
Traditional VQA is significantly impacted by measurement error. JigSaw combats measurement error but introduces high execution cost which is particularly harmful for VQA. VarSaw minimally achieves JigSaw's measurement error mitigation at an execution cost similar to traditional VQA. When measurement error is very high, its benefits in terms of both fidelity and computational cost is even greater.
}
\label{fig:jig_intro}
\end{figure}

\begin{figure*}[t]
\centering
\includegraphics[width=0.9\textwidth,trim={0cm 0cm 0cm 0cm},clip]{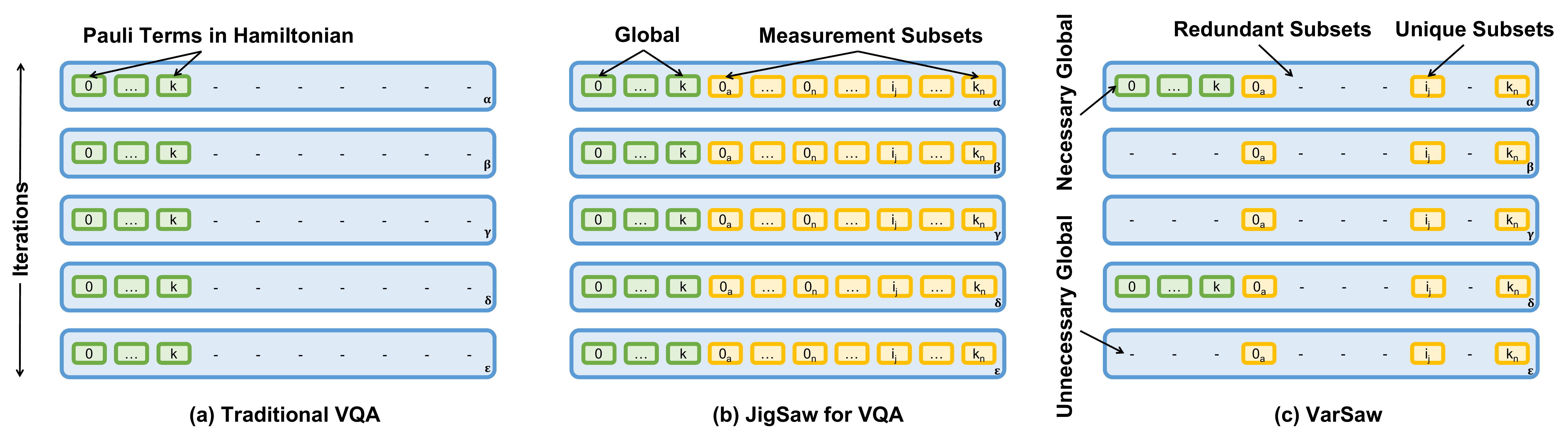}
\caption{
(a) Traditional VQA task run over multiple iterations. Each iteration executes multiple circuits corresponding to the different Pauli Strings in the problem Hamiltonian.
(b) For every circuit in the original task, JigSaw runs multiple circuits - these are the Global circuits (same as original) and the Measurement Subsets (which scale linearly in the number of {circuit qubits). These} are run every iteration.
(c) VarSaw optimizes JigSaw in a VQA-cognizant manner. It exploits spatial redundancy  (repetitions and commutativity) to reduce the number of Subsets executed every iteration. Further, it exploits temporal redundancy (similarity in adjacent iterations' distributions) to reduce the number of iterations on which the Globals are executed.
}
\label{fig:jig_overview}
\end{figure*}


Measurement errors are often the most dominant source of error on current superconducting quantum computers, with average {error rates} ranging as high as 2-7\% \cite{Villalonga_2020} 
Furthermore, measurement operations suffer from measurement crosstalk \cite{jigsaw}
, which means performing several measurement operations concurrently can substantially increase the error rate of each measurement operation by an order of magnitude~\cite{sarovar2020detecting}. 
Thus, larger programs are especially susceptible to measurement errors (crosstalk and otherwise) due to the higher number of measurement operations. 
{This can be particularly detrimental to variational quantum algorithms which a) use low depth circuits and are thus often dominated by measurement errors, and b) can require very high accuracy for real-world {applications such as molecular} chemistry~\cite{cafqa}.}
In this work, we improve the fidelity of VQAs by specifically targeting measurement errors.

Recent work, JigSaw~\cite{jigsaw}, proposed a measurement error mitigation scheme for general quantum circuits.
The key insight in this work  is to divide the target circuit execution into two components.
The first component is the idea of measurement subsetting (the `subset' runs), wherein only a subset of the target circuit's qubits are measured, for all such subsets that span the entire circuit. 
Measurement subsetting encounters fewer errors due to a) appropriately mapping the target logical qubits to be measured onto the physical qubits with highest measurement fidelity, and b) reduced measurement crosstalk, since fewer qubits are measured together. 
But, subsets alone are insufficient to construct the output of the target circuit, since they lack information about the global relations across all the qubits.
The second component involves running the target circuit as is (the `global' run), and then using its outcome to construct a high fidelity, relatively {measurement-error-free output} from the subset runs.
To do so, a Bayesian Reconstruction algorithm is employed, which enables a particular probability distribution (obtained from the subsets) to be appropriately modified using other information (obtained from the Global). Bayesian Reconstruction is inspired by Bayesian updating in statistics, whereby a prior probability estimate is updated using additional information~\cite{joyce2003bayes}.

While this prior work achieves substantial benefits, these benefits come at non-trivial computational cost.
For example, VQE for a relatively small 6-qubit CH$_4$ molecule, when run with JigSaw does not converge to an accurate solution in over 100 hours of VQE simulation.
This is because the additional subset runs scale polynomially (linear to quadratic) in the number of qubits in the circuit. 
While this cost might be tolerable for single-run circuits (the primary focus of their work), the costs are too substantial for variational algorithms, for three reasons: \circled{a}\ VQA is an iterative task, requiring 1000s of iterations to navigate the noisy quantum device landscape to find the optimal solutions \circled{b}\ for real world applications like molecular chemistry, the Hamiltonian is made up of 1000s of Pauli terms/strings, which result in a considerable number of circuits to be executed per VQA iteration; and \circled{c}\ target applications (and thus the quantum circuits) can require 100s of qubits for any potential quantum advantage. Without the additional layer of measurement error mitigation, VQA is already computationally demanding. If used naively with the simplest subsetting, the additional measurement error mitigation layer adds yet another factor of computational requirements by running the same VQA instances multiple times. 
These costs are illustratively quantified and discussed further in Section \ref{prop} and Fig.\ref{fig:jig_motive}.

In this work, we seek to combat the effect of measurement errors in variational algorithms {at a reasonable cost}.
To do so, we build application-aware optimizations to the JigSaw approach and propose \textbf{\emph{VarSaw: Application-tailored Measurement Error Mitigation for  Variational Quantum Algorithms}}.

VarSaw's primary goal is to reduce computational costs by identifying two forms of redundancy: \circled{a}\ \emph{Spatial Redundancy in Subsets:} The generation of measurement Subsets for each Pauli string/circuit of the problem Hamiltonian, leads to significant redundancy (i.e., Subsets can repeat or commute), and \circled{b}\ \emph{Temporal Redundancy in Globals:} The incremental gradient based iterative approach used in VQA optimization leads to significant redundancy among the proximate `Global' runs that JigSaw uses for Bayesian reconstruction (i.e., nearby Globals create nearly the same distributions). Both these forms of redundancy go unnoticed in the prior application-agnostic circuit-focused methodology.

Primarily, VarSaw reduces the quantum resources required to exploit measurement error mitigation by taking an end-to-end, application tailored, holistic approach.
In doing so, it is also able to achieve VQA accuracy improvements.
The same CH$_4$ example from earlier, when run with VarSaw, is able to converge to an accurate solution in under 10 hours of VQE simulation.
An overview of VarSaw is provided in Fig.\ref{fig:jig_overview} and its caption.

\textbf{VarSaw key insights, contributions and results:}
\begin{enumerate}
    \item VarSaw improves JigSaw~\cite{jigsaw}, specifically in the context of VQAs, by identifying \emph{Spatial Redundancy in JigSaw Subsets} (across VQA Pauli strings) and \emph{Temporal Redundancy in JigSaw Globals} (across VQA iterations).
    \item To combat Spatial Redundancy, VarSaw proposes \emph{Commuting of Pauli String Subsets}: which takes a Hamitonian-aware approach to commute and eliminate unnecessary subset circuits generated across all Pauli strings of the problem Hamiltonian. 
    \item To combat Temporal Redundancy, VarSaw proposes \emph{Selective Global Executions}: which takes a VQA tuning-aware approach to only perform Global executions occasionally and increase/decrease their execution via a feedback-based approach. 
    \item In all, VarSaw reduces computational cost over naive JigSaw for VQA by 25x on average and up to 1000x (in terms of number of circuits executed) for the same target accuracy.
    \item Further, it can recover, on average, 45\% of the infidelity from measurement errors in the noisy VQA baseline.
    \item Additionally,  it improves fidelity by 55\%, on average, over JigSaw for a fixed computational budget. 
    \item Importantly, this work showcases the overwhelming benefits from tailoring state-of-the-art optimizations in a domain-specific manner. Rejigging JigSaw from the ground up in a VQA-cognizant manner significantly benefits measurement error mitigation for VQAs, in terms of both computational cost as well as fidelity.
\end{enumerate}


\section{Background}

\subsection{Noisy Intermediate Scale Quantum Computing}
\label{b_NISQ}
Today's quantum devices are noisy~\cite{preskill2018quantum}.
\rev{The sensitivity to external noise channels and imperfections in control and readout circuitry limit qubits from retaining their state indefinitely and causes the error rate of quantum operations to be very high.}
For example, the average error rate of a two-qubit gate is about 1\% on existing hardware from IBM and Google, whereas that of measurement operations is about 4\%~\cite{sycamoredatasheet}, limiting the fidelity of quantum programs on NISQ devices. Consequently, software mitigation of hardware errors on NISQ devices is an active area of research. Multiple forms of error mitigation strategies have been proposed to reduce the impact of different forms of errors.
These include, but are not limited to, noise aware compilation~\cite{murali2019noise,tannu2019not};  correcting measurement errors~\cite{tannu2019mitigating,bravyi2021mitigating}; crosstalk-aware scheduling~\cite{murali2020software,ding2020systematic}; decoherence mitigation~\cite{DDBiercuk_2011,jurcevic2021demonstration,DDKhodjasteh_2007,DDPokharel_2018,souza2012robust, das2021adapt}, spin-echo correction~\cite{hahn}, efficient gate scheduling~\cite{smith2021error}; etc.
In addition, some of these can be used in conjunction to achieve better fidelity~\cite{ravi2021vaqem}.

\begin{figure}[t]
\centering
\includegraphics[width=\columnwidth]{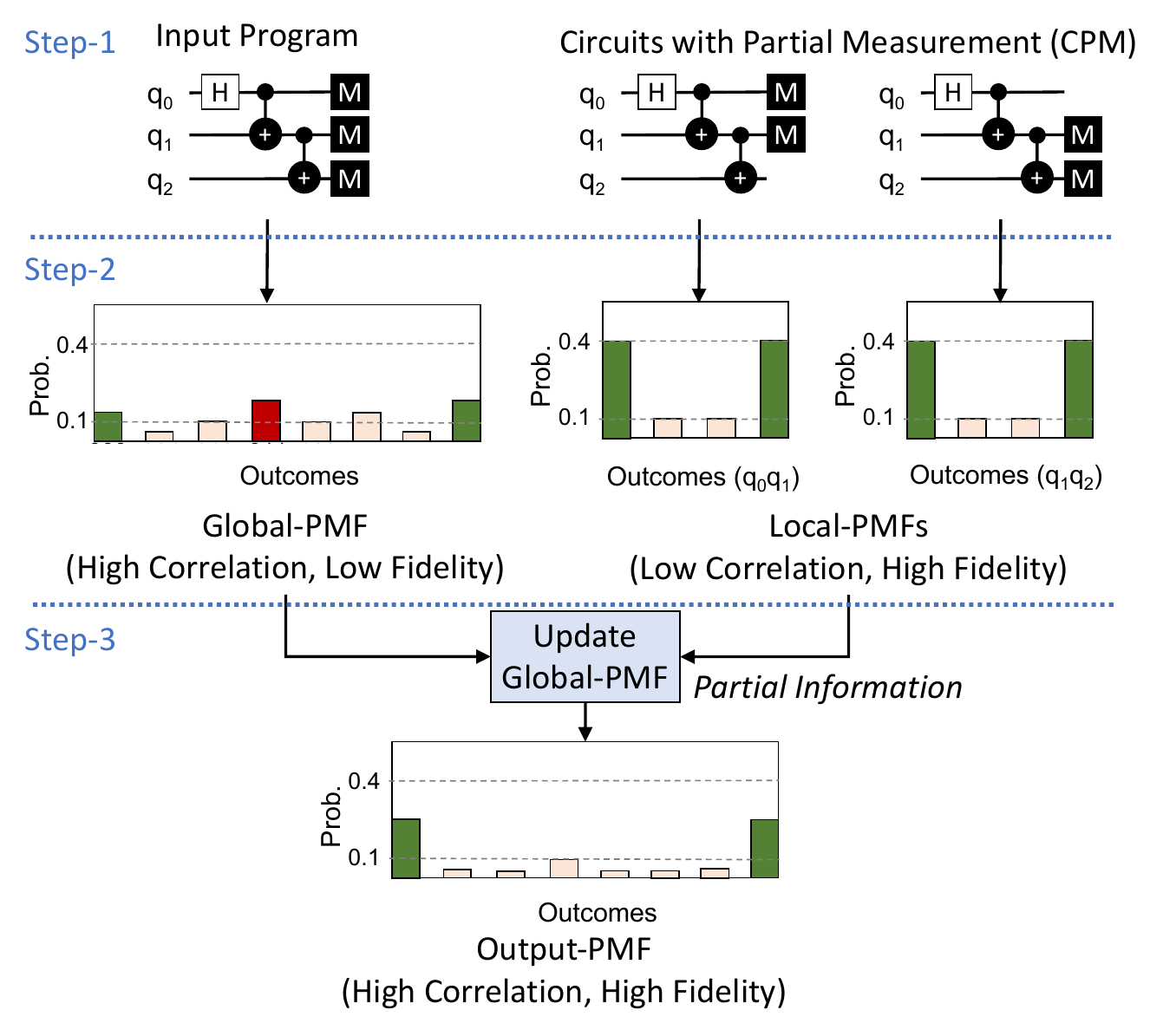}
\caption{
Overview of JigSaw (Adapted from~\cite{jigsaw}).
}
\label{fig:jigsaw_original}
\end{figure}

\subsection{Measurement Errors}
\label{b_Merr}


During computation, the quantum system exists as a linear superposition of many states. To obtain the final outcome, qubits are measured to obtain a classical bitstring. Unfortunately, these operations cannot be implemented perfectly and are a dominant source of error on most existing quantum systems~\cite{sycamoredatasheet}. 
\rev{Program execution is unsuccessful and produces an incorrect outcome even if {one of the measurements fails}. The problem worsens with the program size because the probability of successfully measuring all the qubits simultaneously diminishes rapidly with the number of measurements. Thus, measurement errors constrain the size of the largest program (in terms of number of qubits) that can be run on NISQ devices.} 

Fast and accurate qubit measurements at scale remains an open problem. Measurement operations are imperfect due to several reasons. First, they require coordination between many sophisticated analog and digital instruments 
that operate across thermal domains (from 20 milli-Kelvins to room temperatures)~\cite{krantz2019quantum} increasing susceptibility of these operations to thermal noise. Second, crosstalk arises from the unwanted couplings between the measurement apparatus when multiple qubits are simultaneously measured. For example, simultaneous qubit measurements are 1.26x times more likely to encounter errors on average compared to isolated measurements on Google Sycamore~\cite{sycamoredatasheet}. Last, these operations incur long latencies (about 800ns on Google devices and 4-5 microseconds on IBM's devices\cite{ai2021exponential, jigsaw}) during which qubits may lose their information due to decoherence. It is non-trivial to completely eradicate measurement errors at the device-level~\cite{krantz2019quantum}. Measurement errors manifest as bit-flips and several recent studies specifically focus on minimizing their impact at the application-level~\cite{jigsaw,matrixmeasurementmitigation,bravyi2020mitigating,kwon2020hybrid,FNM,barron2020measurement}.

\subsection{JigSaw for Measurement Error Mitigation}
\label{b_Jigsaw}
JigSaw~\cite{jigsaw} is a recent measurement error-mitigation technique whose overview is shown in Figure~\ref{fig:jigsaw_original} and is comprised of three key steps. First, it takes a program as an input and prepares multiple subset circuits or `Circuits with Partial Measurements' (CPM). Each CPM performs the exact same quantum operations as the input program but measures only a unique subset of qubits. For example, the two CPM in Figure~\ref{fig:jigsaw_original} measures qubit subsets $[q_0,q_1]$ and $[q_1,q_2]$ respectively, for a 3-qubit GHZ program. Second, JigSaw executes the input program and the CPM on the NISQ device. The input program produces a Global-PMF (here PMF stands for Probability Mass Function) that captures the global correlation of the program over all the qubits but has lower fidelity. On the other hand, each CPM produces a partial or marginal Local-PMF over only the measured subset of qubits with high fidelity owing to fewer measurement operations. The third and final step of JigSaw leverages the partial information from the Local-PMFs to adjust the probabilities of each outcome in the Global-PMF using Bayesian reconstruction and {produces a high-fidelity} Output-PMF for the program. By default, JigSaw uses CPM that scales linear in the number of qubits so that high-fidelity Local-PMFs are obtained for each program qubit. In general, for a circuit with $n$ qubits and a subset of size $m$, there exist ${n \choose m}$ CPMs.

{ As an example, for a 3 qubit circuit, if one of the subset circuits targeted the first two qubits, then the subset circuit produces a Local PMF  corresponding to {`00', `01', `10', `11'}. The probability of each of these bitstrings is used to create the 3-qubit Output-PMF. 
For instance, the probability of `00' from the Local PMF is used to create the  probability of `000' and `001' in the Output PMF. This is done by weighing the probability of `00' in the Local PMF  with the relative probabilities of `000' and `001' from the Global PMF. 
The subset circuit suffers from lower measurement errors since it can be mapped to the good qubits, plus it is more sparse and results in lower measurement crosstalk.
On the other hand, Bayesian reconstruction is a statistical method and is hence imperfect and introduces some approximation errors.} 

{A detailed depiction of the reconstruction is illustrated in Fig. 6 of the Jigsaw paper~\cite{jigsaw}. 
Jigsaw performs evaluation up to 18 qubits and shows that even for larger circuits, the benefits from avoiding measurement errors far outweigh errors from Bayesian reconstruction. 
Our own results are in agreement.}

\begin{figure}[t]
\centering
\includegraphics[width=0.8\columnwidth,trim={0cm 0cm 0cm 0cm}]{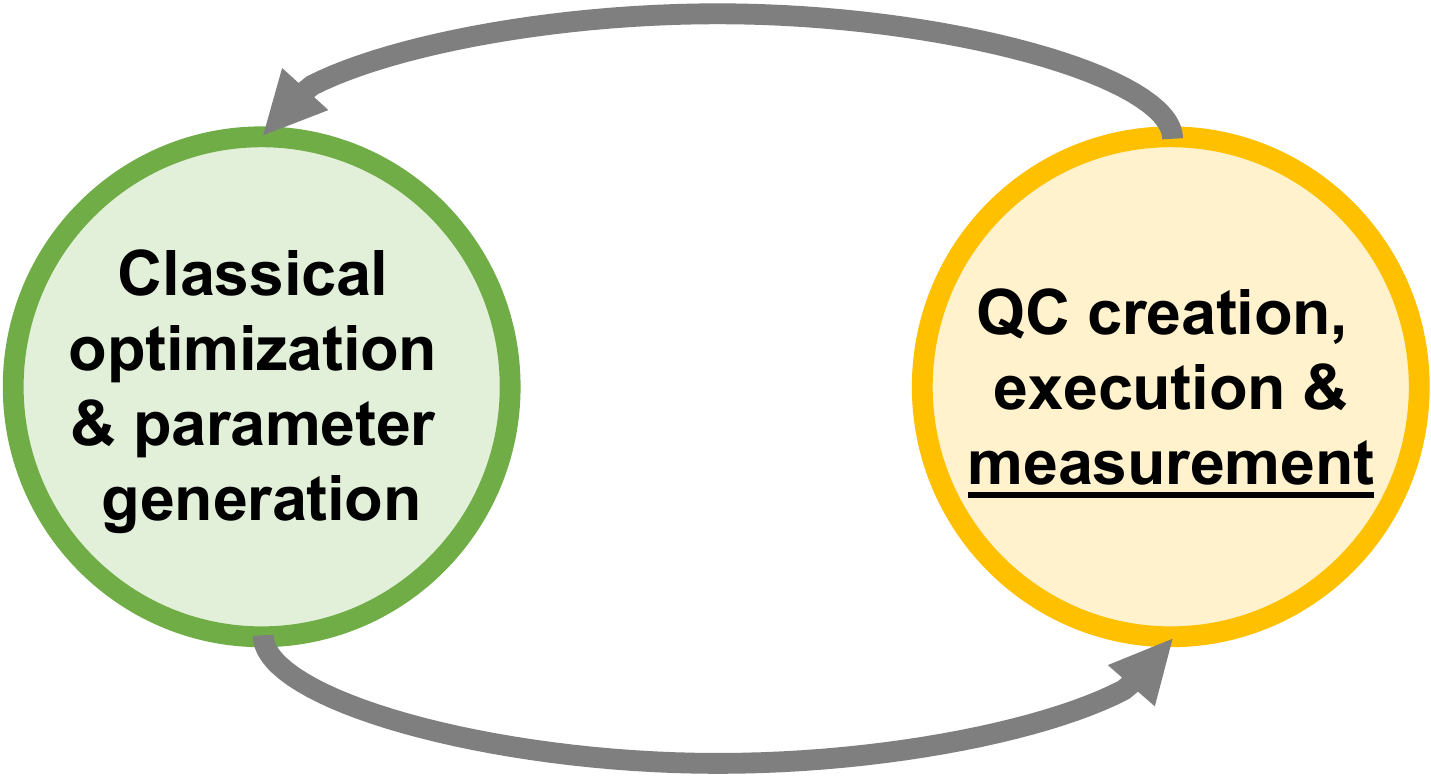}
\caption{VQA: a hybrid algorithm that alternates between classical optimization and quantum execution.}
\label{fig:vqa}
\end{figure}

\subsection{Variational Quantum Algorithms}
\label{b_VQA}
\begin{figure}[t]
\centering
\includegraphics[width=0.98\columnwidth,trim={0.1cm 0cm 0cm 0cm},clip]{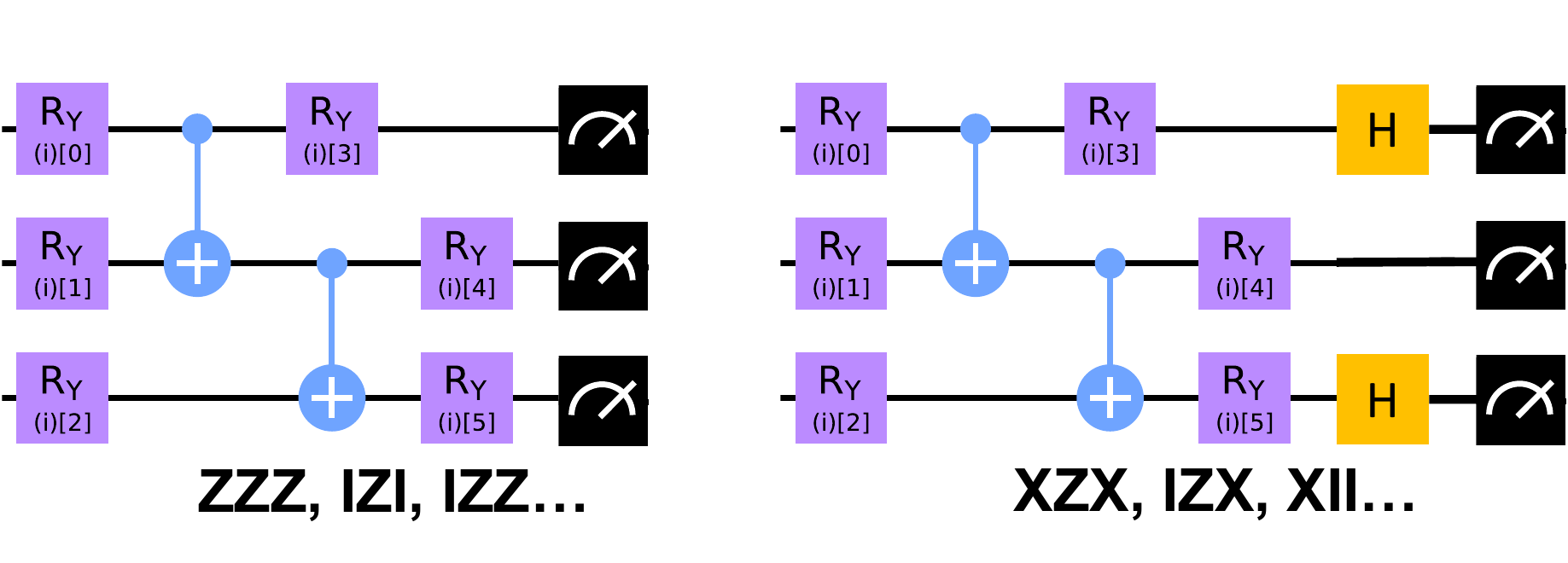}
\caption{
Ansatz circuit measured on different Pauli bases (left: `ZZZ', right: `XZX'). Qubit commutativity allows for a set of Pauli strings to be measured on a single basis.
}
\label{fig:jig_pauli_ansatz}
\end{figure}

Variational algorithms expect to have innate error resilience due to hybrid alternation with a noise-robust classical optimizer~\cite{peruzzo2014variational, mcclean2016theory}. 
An overview of this process is illustrated in Fig.~\ref{fig:vqa}.
There are multiple applications in the VQA domain such as the Quantum Approximate Optimization Algorithm (QAOA)~\cite{farhi2014quantum} and the Variational Quantum Eigensolver (VQE)~\cite{peruzzo2014variational}.
{Our applications in this work target VQE, but VarSaw is applicable to all VQA problems - more on this in Section \ref{FW}.
VQE in itself has wide applications in  molecular chemistry, condensed matter physics, quantum Ising optimization problems, and a variety of quantum mechanics many-body problems, etc.}
An important application of VQE is the ground state energy estimation of a molecule, a task that is exponentially difficult in general for a classical computer~\cite{Gokhale:2019}.
Estimating the molecular ground state has important applications in chemistry, such as determining reaction rates and molecular geometry.
\rev{The quantum circuit used in each iteration of VQE (and VQA in general) is termed an ansatz which describes the range of valid physical systems that can be explored and thus determines the optimization surface. Traditionally, the ansatz is parameterized by 1-qubit rotation gates.
A good ansatz provides a balance between a simple representation, efficient use of available native hardware gates, and sufficient sensitivity with the input parameters.
The classical tuner/optimizer~\cite{9259985,SPSA} variationally updates (often via a gradient based approach) the parameterized circuit until the measured objective converges to a minimum.} 
An example 3-qubit ansatz (i.e., the paramerized circuit which is tuned in VQA) is shown in Fig.\ref{fig:jig_pauli_ansatz} - { measurements have to be} performed on this ansatz every iteration, which is key to the motivation for this work (more on this later).

Estimating the VQE global optimum with high accuracy has proven challenging in the NISQ era even with sophisticated optimizers, a well-chosen ansatz, and error mitigation~\cite{ravi2021vaqem,czarnik2020error,Rosenberg2021,barron2020measurement,botelho2021error,wang2021error,tilly2021variational,takagi2021fundamental}. 
Considering the significant disparity between NISQ VQA accuracy and real world requirements, it is imperative to further aid VQA to the best extent possible.
\begin{table}[t]
    \centering
    \resizebox{\columnwidth}{!}{%
    \begin{tabular}{|c|c|c|c|}
    \hline
         Workload & Ref. Energy & Noisy VQE & VQE+JigSaw \\
         & & & (subset size 2) \\
         \hline
         LiH & 1.72 & 6.11 & 2.73\\
         \hline
         H$_{2}$O & -109.86 & -97.77 & -107.71 \\
         \hline
         H$_{2}$ & 10.46 & 17.00 & 12.51 \\
         \hline
         CH$_{4}$ & -28.55 & -21.22 & -26.63 \\
         \hline
    \end{tabular}
    }
    \caption{{For a given circuit, JigSaw is effective in mitigating measurement errors, improving the circuit's energy estimate.}}
    \label{tab:jigsaw_plus_vqe}
\end{table}

{\subsection{JigSaw for a VQA circuit}\label{b_jigsaw_plus_vqa}}

{As in the case of other circuits, JigSaw successfully mitigates measurement errors at the circuit-level for VQAs.
In Table \ref{tab:jigsaw_plus_vqe}, we show 4 VQE instances from molecular chemistry. In these instances, the VQE ansatz is parameterized with optimal parameters (known from ideal simulation) and executed in noisy simulation.
Clearly, the noisy VQE instance performs considerably worse than the ideal reference simulation. We employ JigSaw on top of this noisy VQE instance and show that JigSaw successfully manages to mitigate measurement errors well, recovering more than 70\% of the optimal energy.
This clearly shows that measurement error plays a significant role in VQA and JigSaw, at a circuit level, is able to achieve considerable benefits.
However, the circuit cost for performing JigSaw is high and the circuit-level benefits can be lost at the application-level, as is discussed next.}


\section{VarSaw: Motivation and Proposal}
\label{prop}

\begin{figure}[t]
\centering
\fbox{
\includegraphics[width=0.95\columnwidth,trim={0.4cm 0cm 1.5cm 0cm},clip]{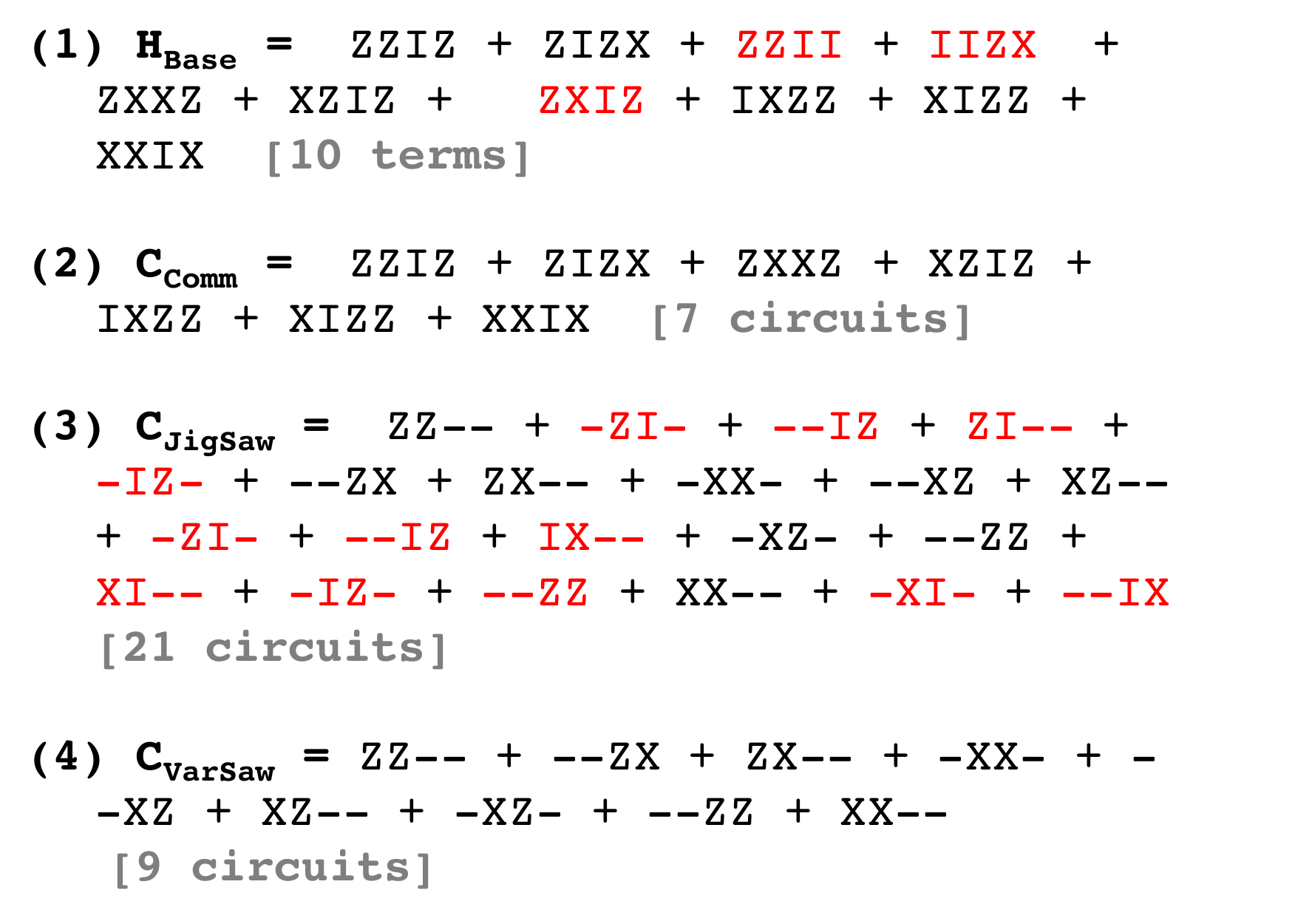}}
\caption{
(1) 4-qubit VQA Hamiltonian with 10 Pauli terms. (2)  Qubit commutativity (red terms in 1) allows for all 10 terms in the Hamiltonian to be measured with  7 circuits/terms. (3) JigSaw uses a 2-qubit sliding window to create measurement Subsets, thus resulting in (4-1)*7=21 circuits. (4) VarSaw identifies redundant / commuting terms in the Jigsaw Subsets (red terms in 3), {to reduce the required} executions to only 9 circuits. 
}
\label{fig:jig_hamil}
\end{figure}

\subsection{Hamiltonian, Paulis, Measurement Commutation}
The VQA problem is represented as a Hamiltonian and is a linear combination of multiple Pauli terms.
The lowest eigenvalue of the Hamiltonian corresponds to the system's ground state energy~\cite{mcclean2016theory}.
Every iteration, the VQA objective function calculates the expectation value of this Hamiltonian.
This objective function is derived from ansatz measurements over different bases.

An example VQE Hamiltonian with 10 terms is shown in Eq.1 in Fig.\ref{fig:jig_hamil}.
This Hamiltonian represents a 4-qubit system, so each term is 4-wide.
While only 10 terms are shown for this trivial example, this number grows quickly for larger Hamiltonians (targeting more complex molecules, for instance).
For example, the $Cr_2$ molecule, which is considered a benchmark for quantum advantage in molecular chemistry minimally requires around 50 qubits and has nearly 50,000 Pauli terms ~\cite{cafqa}.
In general, the number of Pauli strings scales as $O(N^4)$~\cite{gokhale2019minimizing}.
Different Pauli strings are measured by measuring the ansatz in different bases - this is shown in Fig.\ref{fig:jig_pauli_ansatz} for a 3-qubit system.
Different bases correspond to adding appropriate gates at the end of the ansatz.
Not all Pauli strings require different circuits - this is discussed next.

\begin{figure}[h]
\centering
\includegraphics[width=0.75\columnwidth,trim={0cm 0cm 0cm 0cm},clip]{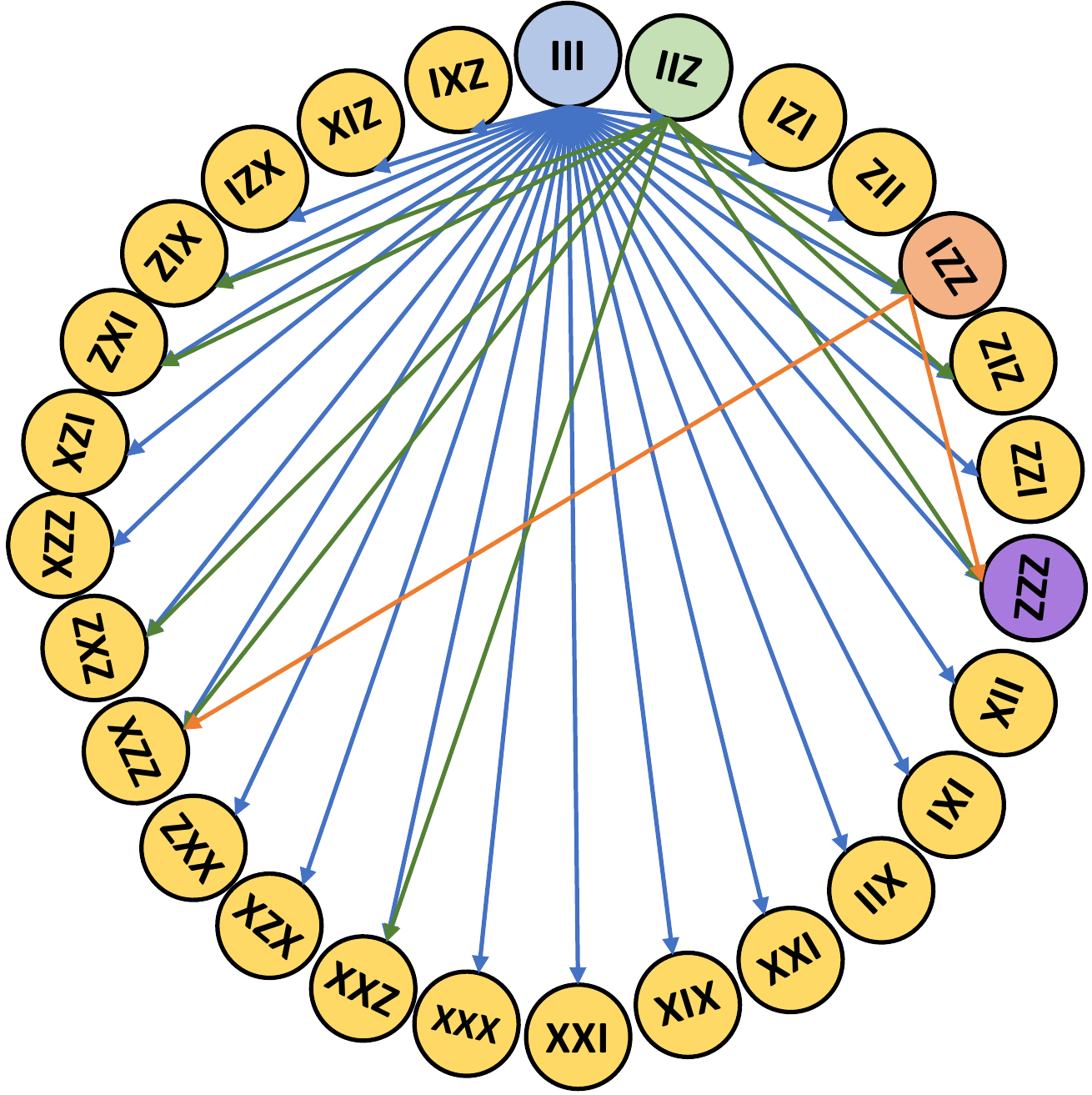}
\caption{
Qubit commutativity graph for  27 3-qubit Pauli strings (`X'/`Z'/`I'). Arrows are drawn from Paulis that can be commutatively measured by others, to the Paulis that can measure them. Arrows are shown for `III' (26), `IIZ' (8) and `IZZ' (2) and `ZZZ' (0).
}
\label{fig:jig_pauli_graph}
\end{figure}

Prior work has shown that the number of circuits executed can be smaller than the total number of Hamiltonian Pauli strings, because strings which correspond to commuting observables can be measured with a single circuit.
The simplest form of this commutation based reduction is shown in Eq.2 in Fig.\ref{fig:jig_hamil} - here the number of terms to be measured reduces from 10 to 7 in Eq.1 since the terms in red trivially commute with one or more terms in black.
A more detailed example of commutation is shown in Fig.\ref{fig:jig_pauli_graph}. 
Here, different Pauli strings (only those made up of `I', `X' and `Z') for a 3-qubit system are shown and the arrows indicate commuting relations.
`III' (blue) commutes with all of the other Paulis (blue arrows) and therefore can use all other Paulis as its commuting parent.
Similarly, `IIZ' has 8 other commuting Pauli parents (green), `IZZ' has 2 (orange) and `ZZZ' has none. 
If the Hamiltonian has the terms `IZZ' and `ZZZ', then a circuit measured in the all-Z basis is sufficient for both of these terms.
Extending from the above, in Fig.\ref{fig:jig_pauli_ansatz} the circuit to the left is used to measure commuting Paulis such as `ZZZ', `IZI', `IZZ' etc.
Other circuits or Paulis can be inferred similarly.

More sophisticated forms of commutation are possible, which can further reduce the number of terms~\cite{gokhale2019minimizing, Shi_2019, hempel2018quantum}, but these forms can: a) non-trivially increase circuit depth and b) can suffer exponential cost to construct.
While VarSaw benefits will potentially be applicable on top of these as well, we limit our scope to the trivial qubit commutations as shown in the examples.

\begin{figure}[h]
\centering
\includegraphics[width=0.98\columnwidth,trim={0cm 0cm 0cm 0cm},clip]{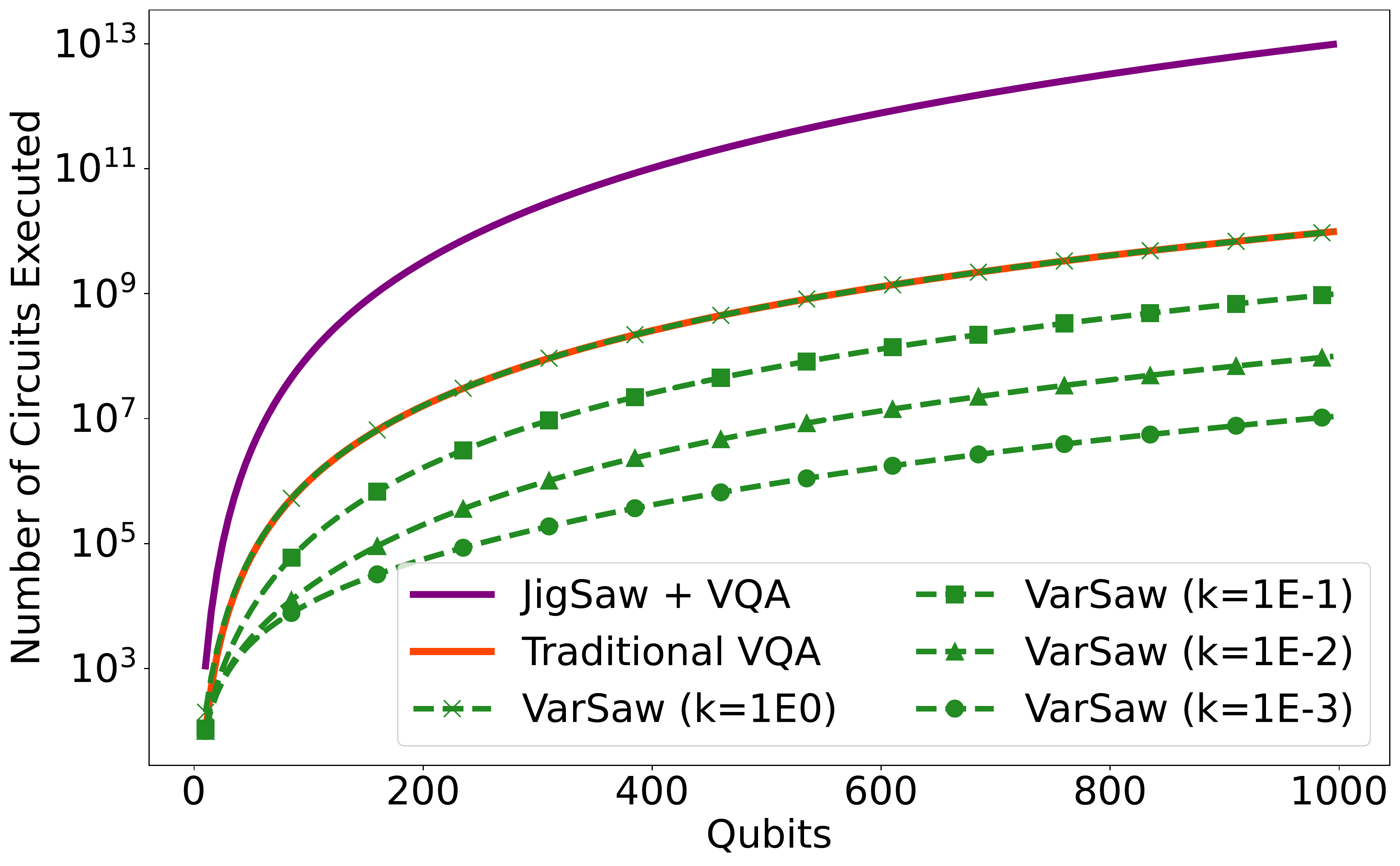}
\caption{
JigSaw incurs high circuit cost for VQA, with executed circuits every iteration roughly scaling as $O(Q^5)$ in the number of qubits,  whereas the baseline scales as $O(Q^4)$. VarSaw scales as $O(Q^2)$ to $O(Q^4)$ while also mitigating errors equal to or better than JigSaw.
}
\label{fig:jig_motive}
\end{figure}

\subsection{Spatial Redundancy in JigSaw Subsets}\label{subsec:spatial_reduction}
Eq.2 in Fig.\ref{fig:jig_hamil} shows that there are 7 Pauli strings post-commutation in $C_{Comm}$.
JigSaw generates Subsets for each of these strings.
The optimal version of JigSaw constructs 2-qubit measurement Subsets using a sliding window, thus 3 subsets are created for each Pauli string, resulting in a total of 21 terms.
This is shown in Eq.3 as $C_{JigSaw}$. 
Now note that there is substantial redundancy among these terms due to repetitions and commutations - these are shown in red. 
For example, `-ZI-', which requires only the 2nd qubit from the left to be measured, commutes with `ZZ--'. 
Other terms can be similarly inferred.
By taking a Hamiltonian-level approach to parsing through the Subsets, VarSaw is able to identify these redundancies, and prunes the number of {terms required to only 9}.
This is shown in Eq.4 as $C_{VarSaw}$.

While the VarSaw term reduction is around 2.3x and the number is equal to $C_{Comm}$ in this example,  more terms or qubits in the Hamiltonian will lead to even greater redundancy, and thus even greater reduction benefits with VarSaw.
This is intuitive from Fig.\ref{fig:jig_pauli_graph}. 
Terms with more `I's have a larger commuting family (i.e., more arrows), thus they are more likely to be combined with other terms.
More generally, the number of terms $H_{Base}$ scale as $O(Q^4)$ \cite{gokhale2019minimizing} where Q is the number of qubits (since each qubit can be `I' / `Z' / `X' / `Y'), whereas the number of terms in $C_{VarSaw}$ scales as $O(Q^{1-2})$ (Sliding window or ${Q \choose 2}$).

This scaling benefit is illustrated in Fig.\ref{fig:jig_motive}.
The graph shows increasing number of circuits executed (on a log scale) against the number of qubits in the problem Hamiltonian \emph{per iteration of VQA}.
The number of circuits for traditional VQA scales roughly as $O(P)$, with roughly $P=0.01*Q^4$, whereas for JigSaw it scales as $O(P + P*Q^{1-2})$, where the first term is the number of global executions and the second is the subsets, with $P$ as the number of Pauli terms in the Hamiltonian. 
On the other hand, the number of terms for VarSaw scales as $O(k*P + Q^{1-2})$, where the first term is the number of global executions and the second is the subsets.
While `k' (0 to 1) and the number of global executions is discussed next, it is evident that VarSaw's scaling is at worst similar to traditional VQA (the line with k=1 overlaps Traditional VQA).
Importantly, VarSaw is at least $O(Q)$ lower quantum computational cost than JigSaw, with at least the same measurement error mitigation benefit.

\subsection{Temporal Redundancy in JigSaw Globals}
\label{TRJG}

\begin{figure}[t]
\centering
\centering \includegraphics[width=0.98\columnwidth,trim={0cm 0cm 0cm 0cm},clip]{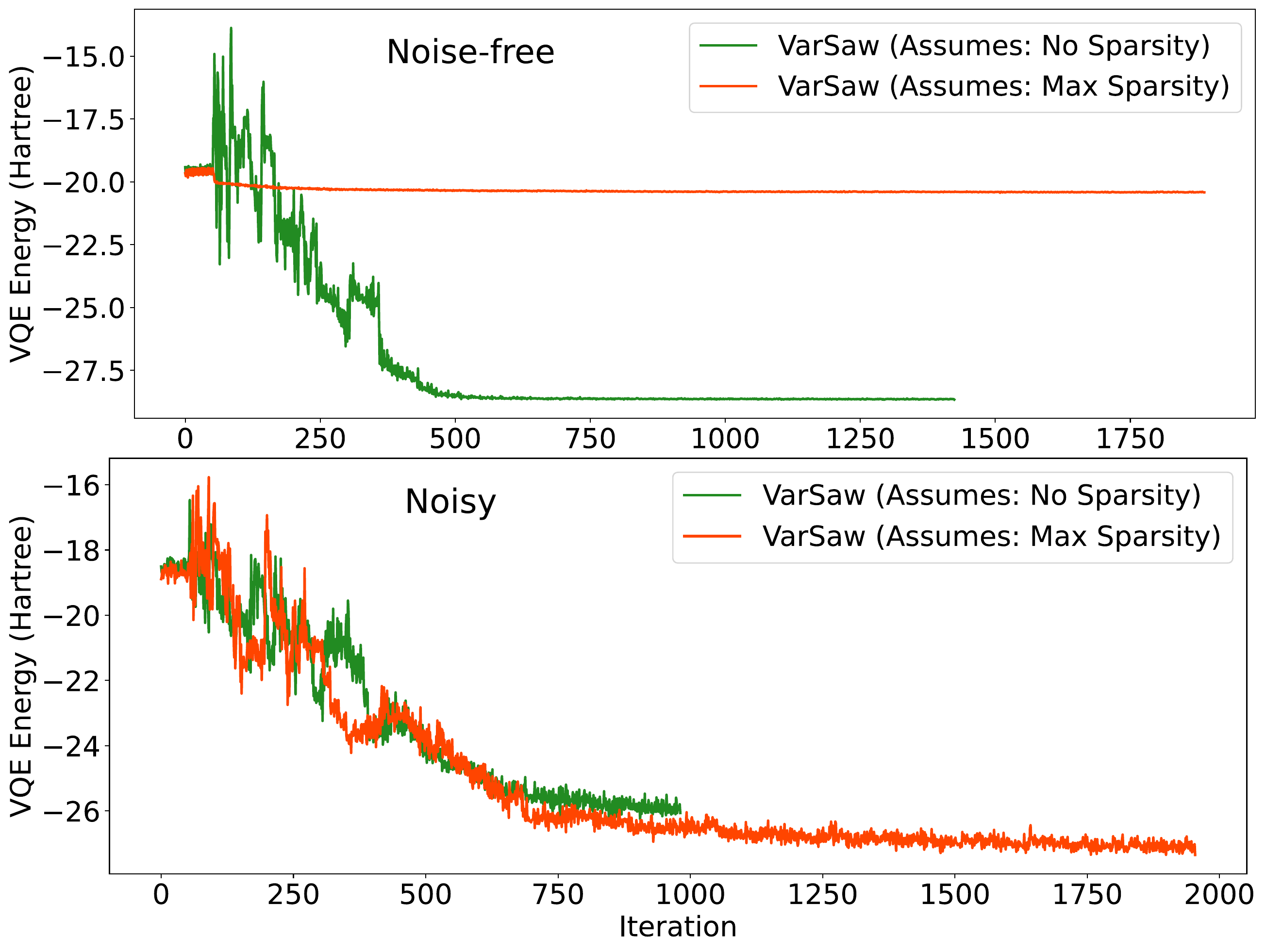}
    \caption{In the absence of noise, increasing VarSaw temporal sparsity in Globals is detrimental to VQA. But in practical settings with considerable noise, temporal sparsity can achieve fidelity similar or better than the alternative, while also achieving considerable reduction in execution cost. }
    \label{fig:jig_temporal}
\end{figure}

Next, we discuss temporal redundancy in JigSaw Globals.
Global executions are performed so as to generate correlations among the different qubits of the target circuit, and these correlations are then used, along with the measurement subsets, to construct the measurement error mitigated probability distribution (via Bayesian Reconstruction).
There are three characteristics of interest associated with the Globals.
First, they suffer from measurement error, since the entire circuit is being executed.
Second, the Bayesian reconstruction itself is associated with some statistical errors.
Third, the Globals are a function of the ansatz parameters of the particular iteration. 

Our hypothesis for temporal redundancy stems from the following:
On the one hand, generating new Globals every iteration will cause lower error due to: a) faithfully following the ansatz parameters, and b) no propagation of Bayesian Reconstruction error across iterations.
On the other hand, generating new Globals every iteration will cause higher error due to even more measurement errors influencing the correlation and reconstruction. 
Thus, as long as the effect of ansatz parameter variation (among proximate iterations) and Bayesian Reconstruction error is minimal compared to the additional measurement errors, it is redundant to perform Global executions every iteration (and potentially beneficial to not do so).
Instead, they can be performed in a more sparse manner, thus further saving computational cost and, in fact, also improving VQA fidelity. 

To analyze the above hypothesis, we construct experiments to compare two extreme scenarios of VarSaw - one performing Globals every iteration (`No-Sparsity'), and the other performing Globals only once, right at the beginning of VQA (`Max-Sparsity').
Both are run for a fixed wall-clock tuning time.
These experiments are performed on a 6-qubit $CH_4$ molecule Hamiltonian with two versions studied in simulation, and are shown in Fig.\ref{fig:jig_temporal}.
The first version is a noise-free experiment with no measurement errors (top).
The second version employs noise mimicking IBM's quantum machine IBMQ Mumbai (bottom).
More details of the experimental methodology is discussed later in Section \ref{methodology}.

In the noise-free experiment, Varsaw with Max-Sparsity performs very poorly compared to the No-Sparsity. 
In the absence of measurement error and any form of significant perturbation, the overwhelming influence of the fixed one-time global execution causes the Max-Sparsity scenario to quickly get stuck at a local minima, whereas the No-Sparsity scenario is faithful to the true distribution and can reach the ideal value.
In the noisy experiment, though, it can be observed that the Max-Sparsity scenario is, in fact, able to marginally outperform the No-Sparsity scenario.
As discussed above, the lower impact of parameter variation and Reconstruction error compared to measurement error, is clearly evident. 
Interestingly, the presence of random perturbation, allows the No-Sparsity scenario to jump out of any early local-minima despite the heavy influence of the fixed single Global execution.
Note that it is not always the case that the Max-Sparsity scenario will outperform  (fidelity-wise) the No-Sparsity scenario, but clearly this is sufficient empirical evidence for the potential similarity in their behaviors.
Also, it is observable in both experiments that the Max-Sparsity scenario is able to perform more iterations than No-Sparsity scenario (for the same time / computational budget). 
This is intuitive, since Max-Sparsity saves on computational resource / time, since it does not run Globals every iteration.
The iteration ratio difference between the top and bottom experiments occurs due to the non-determinism of the tuner, apart from other external variations. 

Perhaps non-intuitively, high sparsity can enable computational cost reduction even over the baseline.
This is because the number of circuits in each global execution or traditional VQA iteration scales as $O(Q^4)$, whereas the number of subset circuits scales as $O(Q^{1-2})$.
The reduced computational cost for different sparsity ratios, as a function of number of qubits, is shown in Fig.\ref{fig:jig_motive}.
Clearly the number of circuits executed is lowered even compared to Traditional VQA as the sparsity of Global executions increases.

It is clear that exploiting temporal redundancy is highly beneficial.
In terms of computational cost, VarSaw clearly benefits over prior work and also potentially over the baseline.
Further, in terms of fidelity, it can potentially achieve even higher fidelity (lower VQA expectation) in comparison to the non-sparse scenarios and over the baseline.

\section{VarSaw Design}

In this Section, we discuss the two design components of VarSaw to exploit and eliminate the spatial and temporal redundancy produced by JigSaw when targeting VQAs.

\begin{figure}[h]
\centering
\centering \includegraphics[width=0.75\columnwidth,trim={9.8cm 0cm 0cm 1.25cm},clip]{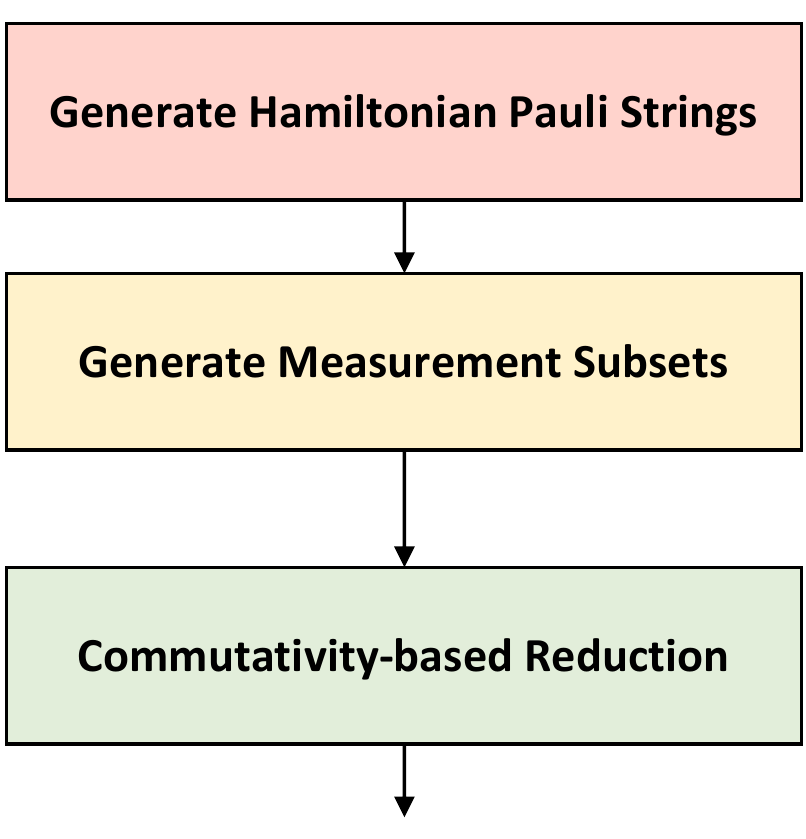}
    \caption{ VarSaw performs subsetting across all Paulis and aggregates them \emph{before} commutativity based reduction.  }
    \label{fig:jig_design1}
\end{figure}

\subsection{Commuting of Pauli String Subsets}\label{subsec:subsetting_design}
The high spatial redundancy in JigSaw's Pauli subsets that repeat and/or commute with each other was motivated in Section \ref{subsec:spatial_reduction}.
The reason for this is that JigSaw is agnostic to the application and is focused on measurement-subsetting only at the circuit level.
VarSaw eliminates this redundancy by integrating measurement-subsetting into the end-end VQA framework.
It performs measurement subsetting immediately after the generation of Hamiltonian Pauli strings and before commutativity-based reduction. 
Once the subsets are generated, commuting terms can then be identified and eliminated, with higher redundancy and thus higher benefits, as was described in Section \ref{subsec:spatial_reduction}. This is shown to the right of Fig.\ref{fig:jig_design1}.

The commutativity based reduction is performed similar to prior work.
For example, ~\cite{kandala2017hardware, smith2016practical} reduce commuting terms on an ad hoc basis, via inspection of the Hamiltonian terms.
More systematic methods are used in OpenFermion~\cite{mcclean2020openfermion} and  Rigetti PyQuil~\cite{computing2019pyquil} libraries.

\begin{figure}[h]
\centering
\fbox{
\centering \includegraphics[width=0.97\columnwidth,trim={0.6cm 0cm 0.5cm 0cm},clip]{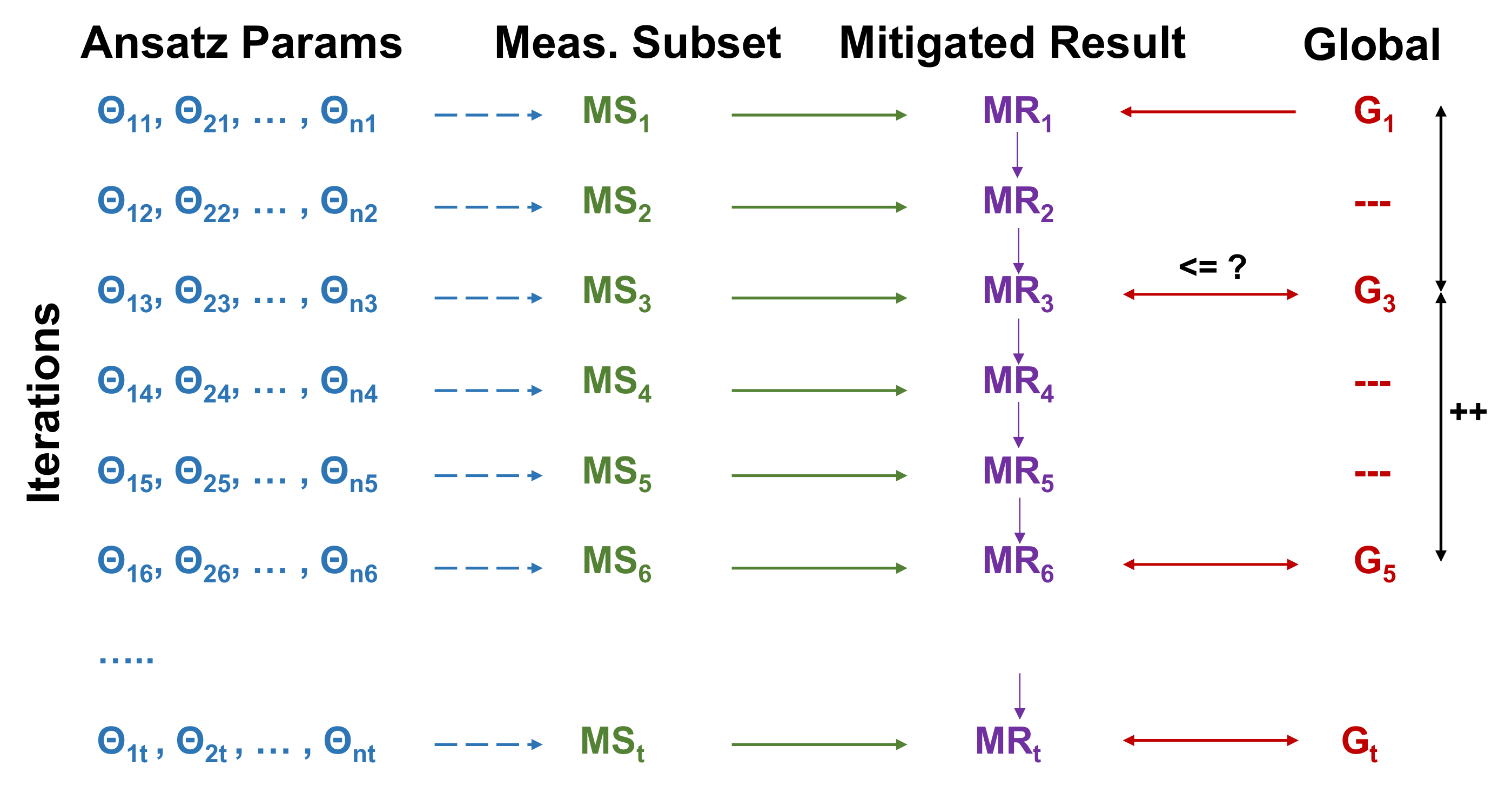}}
    \caption{Design to exploit temporal redundancy by selectively executing the Globals. Measurement subsets are generated every iteration for every new instance of ansatz parameters, whereas the Globals are done only every k iterations where k is dynamically optimized through comparisons against the mitigated results which are derived explicitly from the current iteration's Subsets and implicitly from the most recent Global. }
    \label{fig:jig_design2}
\end{figure}

\subsection{Selective Execution of Globals}

Next, we discuss the VarSaw design to tackle temporal redundancy among the Global executions which, by default, are performed every iteration in JigSaw.
The redundancy exists because, in the presence of non-insignificant noise, the Global execution's influence on the Bayesian reconstruction is less effected by fine granularity ansatz parameter changes and any propagated Bayesian error.
As an additional benefit, by not executing new Globals, there are no new measurement errors introduced. The benefits of reduced measurement errors often overrides the potential loss incurred from eliminating certain Global executions.
Thus, it is sufficient to perform Global Executions sparsely, but how sparse would be dependent on the VQA gradients and the machine noise.

VarSaw designs a simple dynamic optimization scheme to increase or decrease the sparsity of Global Executions depending on its actual impact.
At the beginning, the Globals are set to execute on every $k^{th}$ VQA iteration i.e., these are actual full executions and not just the measurement subsets.
On every non  $k^{th}$ iteration, the Globals are not executed, only the Subsets are run.
The Mitigated Result of this iteration is then generated from the previous iteration's  Mitigated Result and the current iteration's Subsets.
The measurement Mitigated Result for the $k^{th}$ iteration is then computed/verified with the Global and the Subsets --- i.e., we compute the result in both possible ways (a) Executing both Globals and Subsets (b) Executing only the Subsets and using the previous iteration's Mitigated Result. 
If the value of the result in case (a) is less than that in case (b), then the  value of $k$ is decreased.
If not, it is increased. 
This is a simple hill-climbing method, in which the granularity of Globals is increased/decreased by  verifying the quality of the mitigated result.   

This is illustrated in Fig.\ref{fig:jig_design2}.
In the figure, each row represents a VQA iteration, with specific ansatz parameters chosen by the tuner.
The measurement subsets are designed and executed, as discussed in Section \ref{subsec:subsetting_design}.
On iteration 1, a Global execution is performed, and the initial sparsity window is 2 cycles (i.e., the Global execution is set to perform every alternate cycle).
The mitigated result $MR_1$ is then generated from $MS_1$ and $G_1$.
On iteration 2, there is no $G_2$, only $MS_2$. 
Instead $MR_2$ is generated from $MS_2$ and $MR_1$.
On iteration 3, $MR_3$ is still generated from $MR_2$ and $MS_3$. But $G_3$ is also run and the result obtained from $G_3$ and $MS_3$ is compared against $MR_3$. 
If $MR_3$ has similar or lower energy than the result computed using $G_3$ and $MS_3$, that means that the error from fine-granularity ansatz parameter change and propagated Bayesian error is no worse than additional measurement errors.
Thus, $MR_3$ is kept as is, and Global executions are made more sparse.
In the figure, the next Global execution is set to the iteration 5 - doubling the sparsity.
If $MR_i$ is greater than the result obtained using $G_i$ and $MS_i$, then $MR_i$ is instead regenerated with $G_i$ and the sparsity is lowered.

\section{Methodology}\label{methodology}
\begin{table}[]
\centering
\resizebox{0.85\columnwidth}{!}{%
\begin{tabular}{|l|r|r|l|}
\hline
\textbf{Molecule} & \multicolumn{1}{l|}{\textbf{Qubits}} & \multicolumn{1}{l|}{\textbf{Pauli terms}} & \textbf{Temporal?} \\ \hline
$H_2$   & 4  & 15    & Y \\ \hline
$LiH$  & 6  & 118   & Y \\ \hline
$LiH$  & 8  & 193   & Y \\ \hline
$H_2O$  & 6  & 62    & Y \\ \hline
$H_2O$  & 8  & 193   & Y \\ \hline
$H_2O$  & 12 & 670   & N \\ \hline
$CH_4$  & 6  & 94    & Y \\ \hline
$CH_4$  & 8  & 241   & Y \\ \hline
$H_6$   & 10 & 919   & N \\ \hline
$BeH_2$ & 12 & 670   & N \\ \hline
$N_2$   & 12 & 660   & N \\ \hline
$C_2H_4$ & 20 & 10510 & N \\ \hline
$Cr_2$  & 34 & 32699 & N \\ \hline
\end{tabular}%
}
\caption{Molecules for VQE, for VarSaw evaluation, with Qubits and Pauli terms in the Hamiltonian. `Temporal?' indicates if evaluation of Temporal Redundancy is performed.}
\label{tab:Apps}
\end{table}

\begin{figure}[t] 
\centering
\centering \includegraphics[width=\columnwidth,trim={0cm 0cm 0cm 0cm},clip]{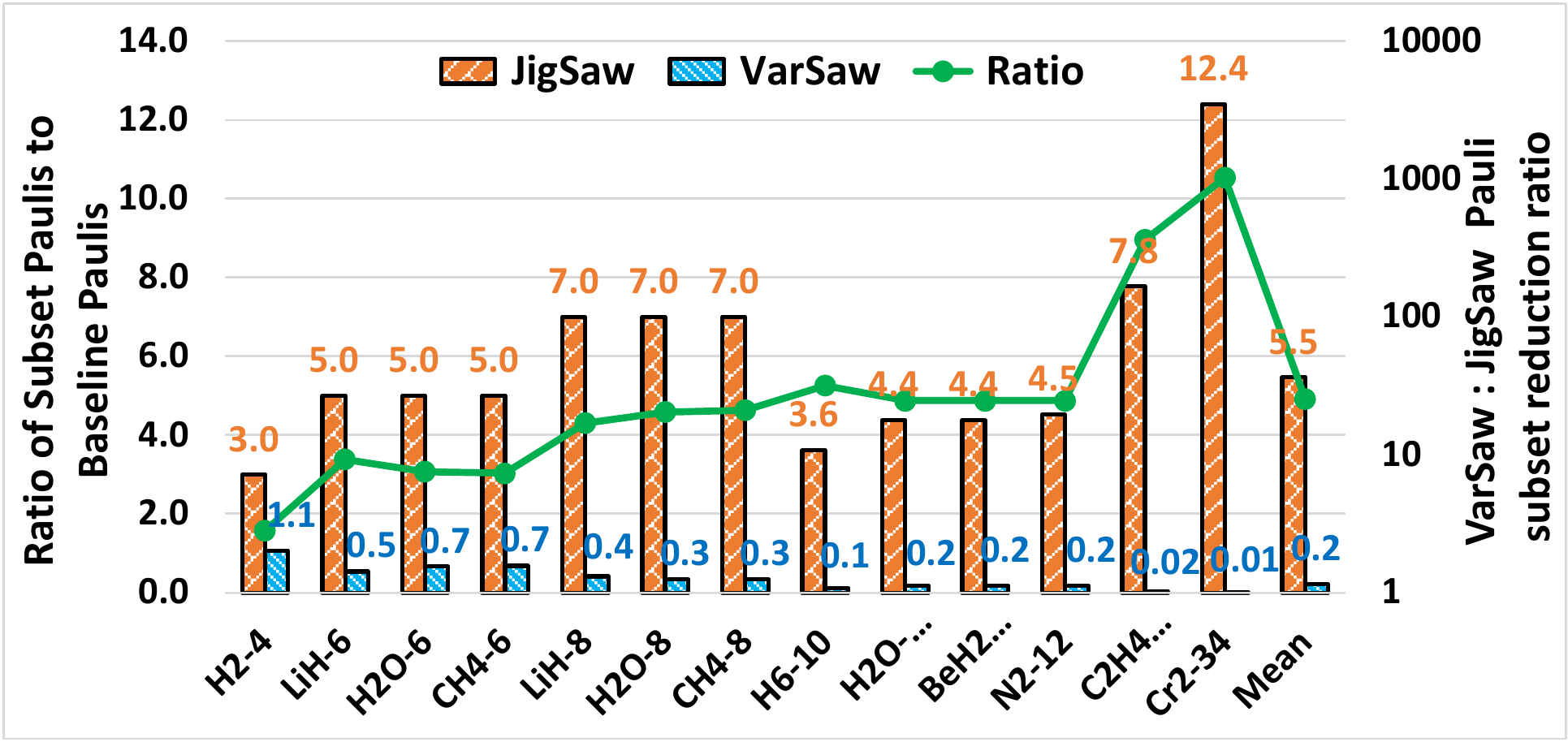}
    \caption{Pauli term reduction in Measurement Subsets achieved by VarSaw compared to JigSaw. Orange columns / left axis: VarSaw and JigSaw subsets expressed relative to the number of Pauli terms in the baseline. Green line / right secondary axis: reduction in  subsets for VarSaw rel. to JigSaw. }
    \label{fig:jig_eval1}
\end{figure}

\subsection{Applications}
We limit ourselves to one VQA domain, the VQE, which was introduced in Section \ref{b_VQA}.
Our evaluations encompass 9 different molecules, the details for which are provided in Table \ref{tab:Apps}.
Some molecules are run with more than a single molecular configuration. Systems with more molecular orbitals frozen require less qubits. The ideal noiseless ground state energy of a molecule is the same across different specifications.

Note that only a subset of the applications are used for full VarSaw evaluation of temporal+spatial redundancy elimination.
This is because the larger molecules/configurations have too many terms/qubits to effectively simulate VQA (especially since we need to run 100s of VQA iterations). For these, we only show the spatial benefits of VarSaw. Applications where we demonstrate spatial+temporal redundancy elimination are restricted to systems of up to 8 qubits.

We use the hardware efficient SU2 ansatz which are of low depth and therefore suited to NISQ devices. 
The ansatz is constructed for the `full' entanglement, which means that entanglement (i.e. $CX$ gates) is allowed between all qubit pairs. 
2 blocks of repetition are used in the ansatz, where each block is an additional layer of $CX$ gates and tunable 1-qubit gate parameters. 
We use the SPSA~\cite{SPSA} and ImFil~\cite{9259985} classical tuners across all our evaluations. 
We find a subset size of two qubits to be optimal to our work, fairly in agreement with JigSaw (evaluation in Appendix.\ref{eval5})

\subsection{Infrastructure and Optimization Process}
We integrate VarSaw as part of the Qiskit~\cite{Qiskit} framework.
The tackling of spatial redundancy is implemented as a pre-processing step before actual VQA execution - this was illustrated in Fig.\ref{fig:jig_design1}. 
Qiskit interfaces with the PySCF library~\cite{sun2017pythonbased} in the process of constructing Hamiltonians from molecular specifications.
JigSaw is then performed after Hamiltonian generation.
This is followed by the commutativity step to reduce the number of terms.

Varsaw's tackling of temporal redundancy of Jigsaw Globals is an on-the-fly  optimization that can be integrated with any VQA classical optimizer. 
We implement it in Python and use it with the Qiskit VQA framework, through which it interacts with quantum execution.

Due to the large number of VQA Hamiltonian Pauli terms, as well as the number of VQA iterations, we primarily evaluate VarSaw on noisy simulation modeled on IBM Quantum Devices \rev{(except in Section \ref{real-device} which shows real machine analysis)}.
We use the noise model of IBMQ Mumbai (27 qubits) --- machine details can be found on the IBM Quantum Systems page~\cite{IBMQS}. 
While Qiskit Runtime~\cite{qiskitruntime} offers support to run iterative applications on the actual quantum hardware, it is very limited in its support for on-the-fly optimizations, hence unsuited to evaluate VarSaw.
Running one iteration at a time on the quantum hardware without Runtime (i.e., using our own classical computer between every quantum execution) is possible, but impractical for most long running applications since machine availability is very sparse and wait-times between iterations are enormous (hours to days).
{Where applicable, results are averaged over up to 10 different trials which run the VQA optimizer with different random seeds.}

\subsection{Evaluation Comparisons and Metrics}

\noindent \textbf{Comparisons:}

\noindent \emph{Baseline:} Traditional VQA without measurement error mitigation, but using Pauli string commutation.

\noindent \emph{JigSaw:} JigSaw applied on top of the Baseline, with each VQA iteration employing 2-qubit measurement Subsets using a sliding window, along with Global executions.

\noindent \emph{VarSaw:} The proposed approach which optimizes JigSaw (for VQA) for spatial and temporal redundancy. 

\vspace{1mm}
\noindent \textbf{Metrics:}

\noindent \emph{Cost:} Quantum computational cost, in terms of the number of quantum circuits that have to be executed on the quantum device (Lower is better).

\noindent \emph{Energy:}  Molecular VQE ground state energy estimates using the standard Hartree Energy metric (Lower is better). This is the equivalent of fidelity for VQA.

\section{Evaluation}

\subsection{Computational cost reduction in Subsets}

Fig.\ref{fig:jig_eval1} shows the Pauli term reduction in Measurement Subsets achieved by VarSaw in comparison to JigSaw. 
Both VarSaw and JigSaw subsets are expressed relative to the number of Pauli terms in the baseline in the orange columns / left axis.
The number of Pauli terms/circuits in JigSaw increases loosely in proportion to the number of qubits.
The worst case increase would be $(Q-1)*P$ where Q is the number of qubits and P is the number of baseline Pauli strings.
The number is often lower than this because measurement subsets which have are all `I's are already weeded out i.e, those terms will require no measurements.
But clearly, the large increase in the number of terms is evident.
Our largest molecule, a 34-qubit $Cr_2$ molecule, sees a 12x increase in the number of terms,
while an average increase of 5.5x is observed for molecular systems ranging from 4-qubits to 34-qubits.

On the other hand, the Pauli term Measurement Subsets in VarSaw is substantially lower due to the elimination of redundant repetitions and commuting terms. 
On average, the number of subsets is only 0.2x of the total Pauli terms.
Thus, the maximum per-iteration computational cost increase for VarSaw is, on average, only 20\% greater than the baseline, while achieving the benefits of measurement error mitigation. 
Note that the cost-per-iteration incurred by VarSaw is usually much less than the baseline since for most iterations, the Globals are not executed at all. 
This is discussed later in Section \ref{eval3}.

Also notable is that the number of Subsets, relative to the baseline Pauli terms, decreases for larger molecules.
This is because, as the number of baseline Pauli terms increases, there is a considerably larger fraction of the Subsets which repeat or commute.
This is especially promising for the scalability of VarSaw's benefits.
For example, $Cr_2$ only requires a 1\% increase in the worst-case computational cost.

On the green line graph / right secondary axis, we show the reduction in measurement subsets for VarSaw compared to JigSaw.
Note that the axis is on a log scale.
Term reduction is greater than 1000x for the largest molecule, while the average reduction is around 25x.
This clearly highlights the benefits of VarSaw term reduction compared to JigSaw, and in general is indicative of the potential of application-aware improvements to state-of-the-art quantum optimizations.

\rev{Overall, VarSaw achieves a 25x lower runtime on average compared to JigSaw, and as much as 1000x lower runtime for our largest problems.
Further, VarSaw has a 10x lower runtime on average compared to the baseline.}

\begin{figure}[t]
\centering
\centering \includegraphics[width=\columnwidth,trim={0cm 0.5cm 0cm 0cm},clip]{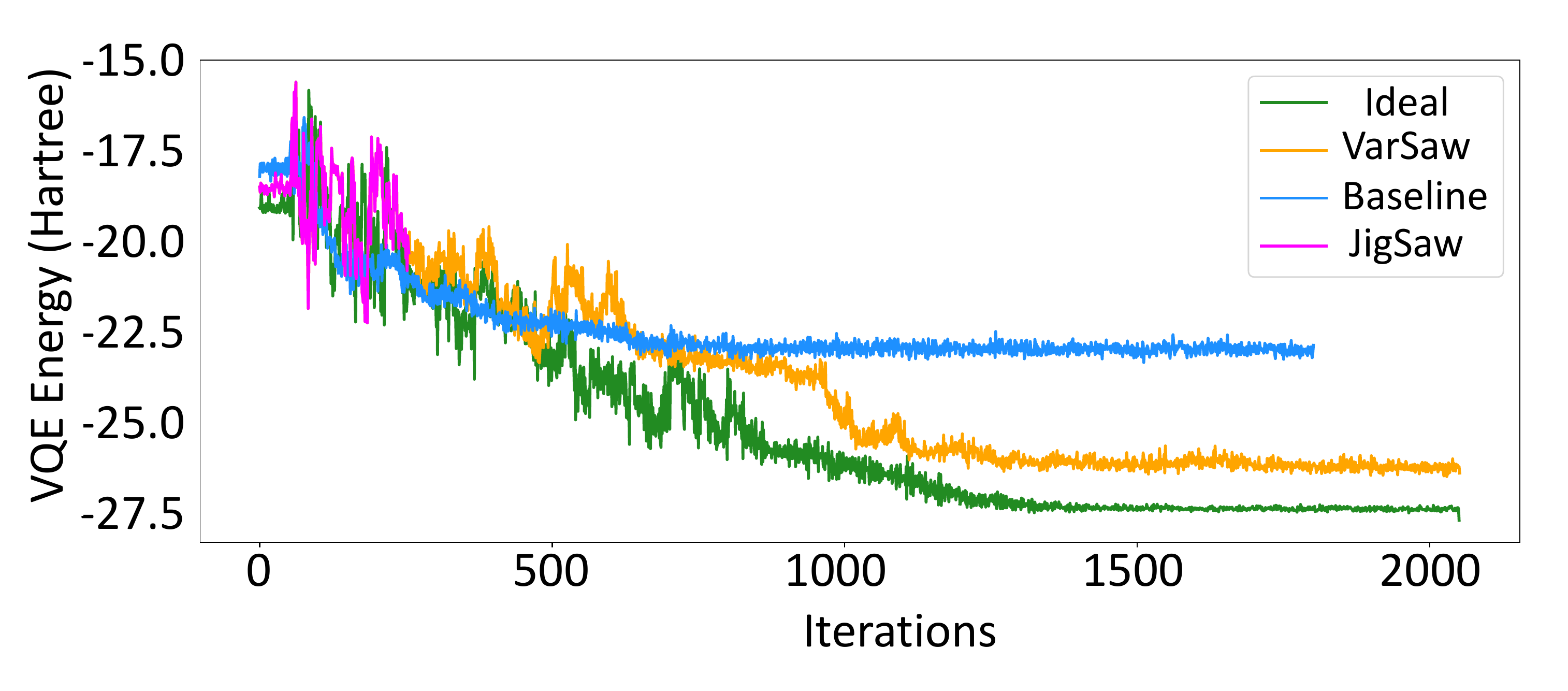}
    \caption{VQE accuracy improvements by VarSaw compared to the noisy baseline and noise-free ideal for CH$_4$ }
    \label{fig:jig_eval2_app}
\end{figure}

\subsection{Single-application $CH_4$ analysis}
Fig.\ref{fig:jig_eval2_app} shows the VQE energy plot for VarSaw, compared to JigSaw, the noisy baseline, and the noise-free ideal, for the CH4 molecule. 
These results are shown for a fixed circuit budget, i.e., all 4 scenarios run the same number of circuits.
VarSaw is able to achieve the highest accuracy of VQE energy estimates, comparable to the Ideal.
The baseline is affected by measurement errors which is expected.
It also runs fewer iterations than VarSaw.
This is because VarSaw has a lower cost per iteration due to minimal number of global iterations.
This is discussed further in Section \ref{eval3}.
On the other hand, while JigSaw also performs measurement error mitigation, it is only able to run a fraction of the iterations due to its high circuit cost per iteration. 
Thus, it achieves a lower accuracy than the baseline in this fixed budget scenario. 
VarSaw to JigSaw accuracy comparison, under a fixed circuit budget, is discussed further in Section \ref{eval4}.

While other forms of error continue to persist (and is orthogonal to this work), it is evident that the subsetting technique is able to eliminate a considerable fraction of the measurement error, leading to significant accuracy improvements. 
Further, VarSaw is able to do so at low computational cost, meaning that it is able to optimize further, to better accuracy, under a fixed computational budget. 

\begin{figure}[t]
\centering
\centering \includegraphics[width=\columnwidth,trim={0cm 0cm 0cm 0cm},clip]{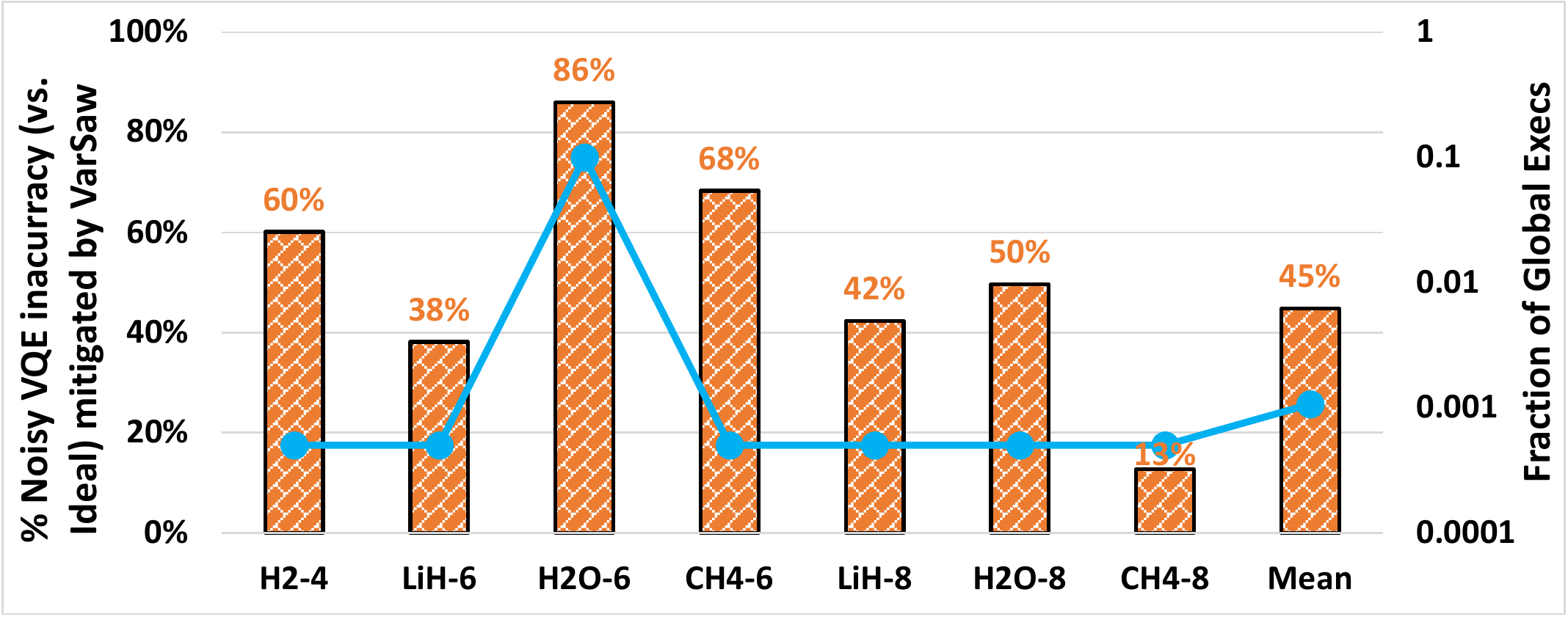}
    \caption{VQE accuracy improvements by VarSaw compared to the baseline over 2000 iterations. Secondary axis shows the optimal fraction of global executions  required for VarSaw. }
    \label{fig:jig_eval2}
\end{figure}

\subsection{VQE Accuracy Improvements through Error Mitigation and Global execution fraction}
\label{eval3}
Next, we show the improvements in VQE accuracy achieved by VarSaw compared to the baseline over 2000 iterations of VQA tuning in Fig.\ref{fig:jig_eval2} (orange columns / left axis). 
The improvements are achieved thanks to the measurement error mitigation adopted from the JigSaw approach.
VQE accuracy improvement ranges from 13\% to 86\% with a mean improvement of 45\%.
Note that the results shown here are only for the smaller benchmarks (up to 8 qubits) since the larger ones are not amenable to noisy simulation. 
Clearly, VarSaw achieves considerable improvement to VQE accuracy while only resulting in minimal increase to the worst case per-iteration computational cost. 


The blue line / right axis shows the optimal fraction of global executions required for VarSaw.
Clearly the fraction is very low, meaning that performing only a few Globals is most beneficial.
This is so for two reasons.
First, performing fewer Globals means that new measurement errors are not being introduced into the tuning process.
Second, fewer Global executions means that the computational resources (i.e., quantum circuits) can be utilized towards running more subset iterations - allowing for further VQA optimizations, albeit with imperfect correlation.
VarSaw performs optimally at a Global granularity of around  1 in 100 iterations, which means that the total number of circuits executed, in comparison to the baseline, is also more than 10x lower. 
This can, of course, change with different noise characteristics, problem complexity etc, but VarSaw will always be computationally competitive with the baseline while achieving significant measurement error mitigation benefits. We evaluate VarSaw's benefits over a noise sweep in Appendix.\ref{eval6}.

\begin{figure}
\centering
\centering \includegraphics[width=\columnwidth,trim={0cm 0cm 0cm 0cm},clip]{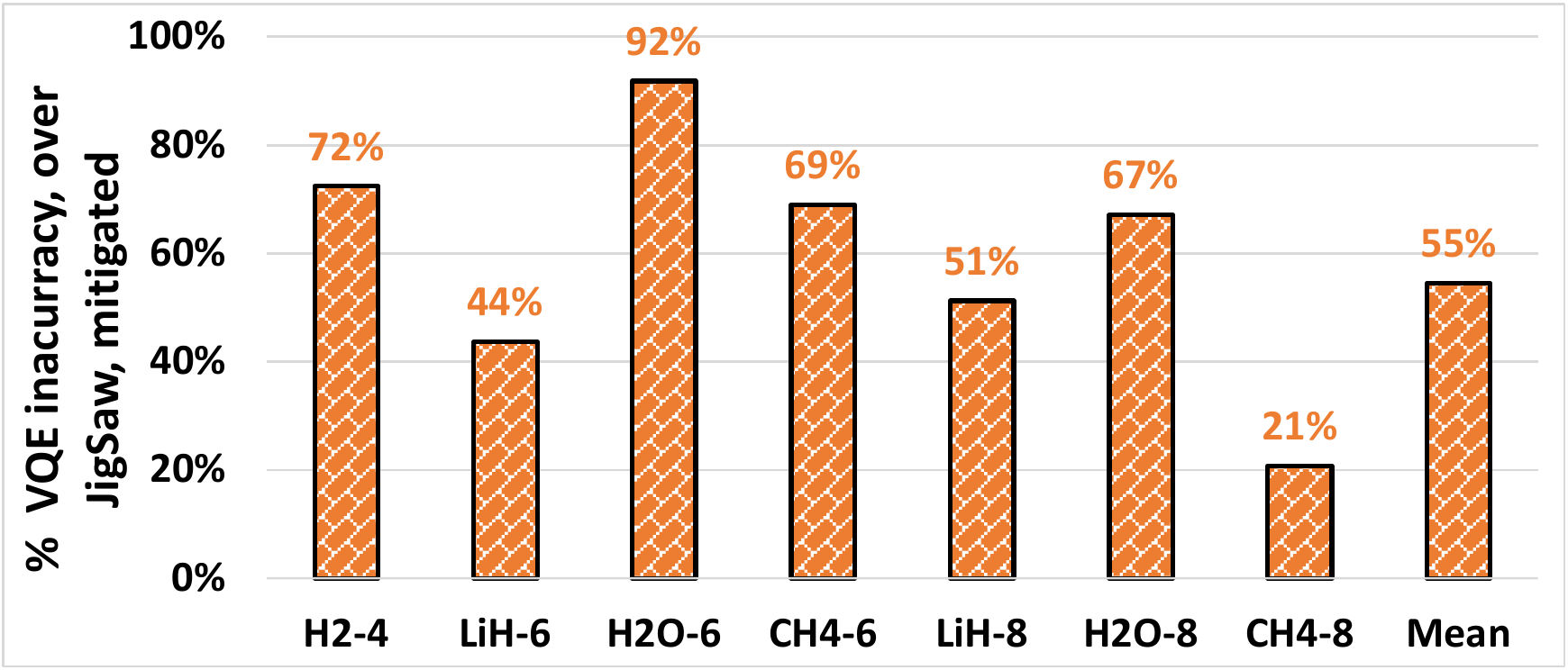}
    \caption{VQE Accuracy Improvements over JigSaw for fixed circuit budget }
    \label{fig:jig_eval3}
\end{figure}

\subsection{VQE Benefits over JigSaw for fixed circuit budget}
\label{eval4}
While VarSaw employs the same error mitigation as JigSaw, it is much more computationally efficient.
Thus, under a fixed circuit execution budget, the VQE accuracy improvements achieved by VarSaw is considerably greater than that of JigSaw.
This is because VarSaw can execute orders of magnitude greater number of VQA iterations for the same circuit budget.
This is important both in terms of quantum device availability in the cloud as well as monetary cost which scales proportionally to the circuits executed.  
Fig.\ref{fig:jig_eval3} shows the improvements in VQE accuracy achieved by VarSaw compared to JigSaw for a fixed circuit budget that allows VarSaw to complete nearly 2000 iterations, but JigSaw can only complete a few 100s.
Thus, VQE accuracy improvement for VarSaw over JigSaw ranges from 21\% to 92\% with a mean improvement of 55\%.
These benefits will increase as we scale to larger molecules since the computational cost difference between JigSaw and VarSaw increases with system size as discussed earlier.

\begin{figure}[h]
    \centering
    \includegraphics[width = \columnwidth]{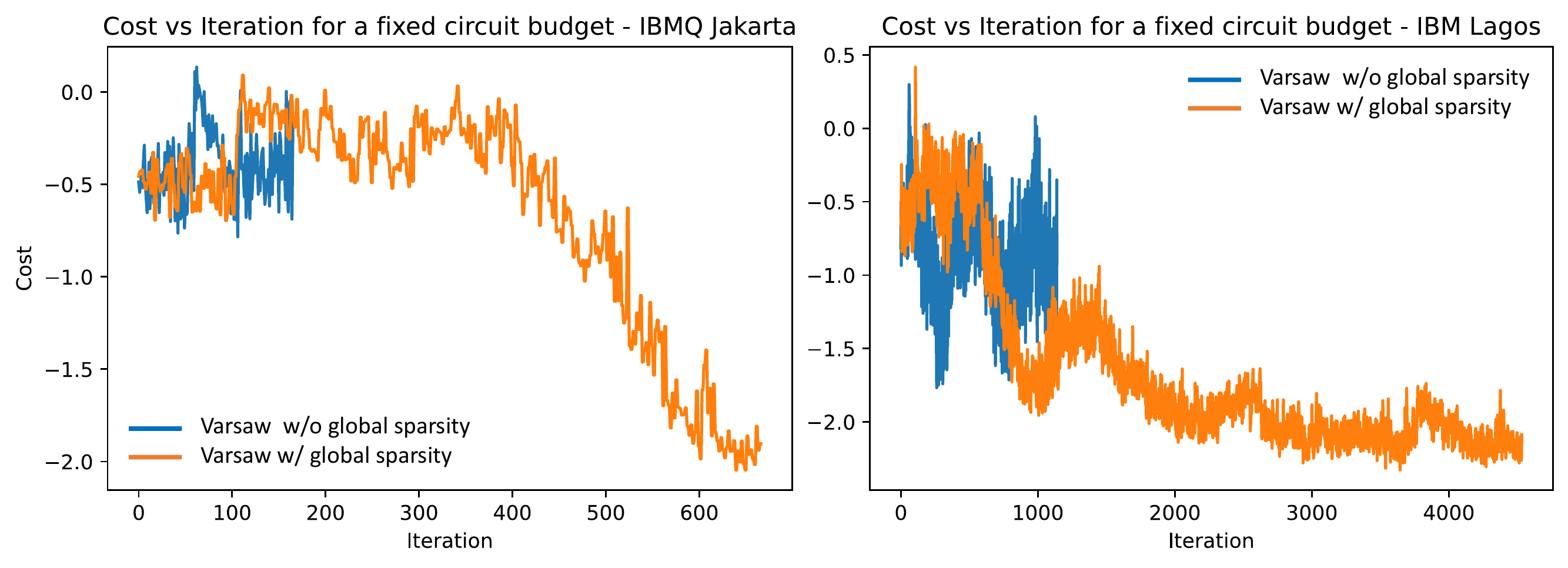}
    \caption{\rev{Varsaw on real IBM Q devices - Lagos and Jakarta}}
    \label{fig:varsaw_device}
\end{figure}

\subsection{\rev{Real device experiments}}
\label{real-device}

\rev{
The experiments so far have focused on noisy simulated quantum devices. 
In this section, we discuss smaller VQA experiments run on real quantum devices (IBM's Lagos and Jakarta) to highlight the benefits of VarSaw's temporal optimization by selectively executing the Globals.
This is shown in Fig.\ref{fig:varsaw_device}.
To keep circuit overheads to a minimum (for the runs to be amenable to real device execution), we run VQE on a Transverse Field Ising Model (TFIM) Hamiltonian with 5 qubits and only 3 Pauli terms.}

\rev{It is clearly evident that the VarSaw temporal optimization that employs the Global selective execution / sparsity, considerably reduces the circuit overhead per VQA iteration while still being faithful to the overall Hamiltonian objective. Therefore, sparse VarSaw is able to complete a significantly greater number of meaningful VQA iterations (nearly 4x) compared to Varsaw with full Global execution, and achieves 1.5x-3x VQA objective improvements  on real devices. These benefits are expected to be greater for more complex VQA problems which will have even greater impact of measurement fidelity and will thus gain from the VarSaw method.}

\rev{Finally, note that for the VarSaw spatial optimization, the benefit is orthogonal to running on a real device since the primary benefit is circuit overhead reduction which is the same for the real device or simulation.}

\subsection{\rev{VarSaw benefits on different ansatz structures}}

\begin{table}[]
    \centering
    \begin{tabular}{|c|c|c|c|c|}
    \hline
         Workload & Full & Linear & Circular & Asymmetric  \\
         \hline
         $CH_{4}$ & 44.42 & 65.87 & 23.26 & 31.24 \\
         \hline
         $H_{2}O$ & 32.09 & 96.49 & 63.86 & 54.51 \\
         \hline
         $LiH$ & 26.90 & 85.09 & 37.86 & 65.64\\
         \hline
         
    \end{tabular}
    \caption{\% VQE Inaccuracy Mitigated by VarSaw with Global Selective Execution, over VarSaw without Global Selective Execution, for Different Ansatz Types}
    \label{tab:ansatz_type}
\end{table}

\begin{table}[]
    \centering
    \begin{tabular}{|c|c|c|c|c|}
    \hline
         Workload & p = 1 & p = 2 & p = 4 & p = 8\\
         \hline
         $CH_{4}$ & 29.98 & 26.73 & 21.17 & 5.29\\
         \hline
         $H_{2}O$ & -1.46 & 31.83 & 27.88 & 7.97 \\
         \hline
         $LiH$ & 58.67 & 53.59 & 26.04 & 7.79\\
         \hline 
    \end{tabular}
    \caption{\% VQE Inaccuracy Mitigated by VarSaw with Global Selective Execution, over VarSaw without Global Selective Execution, for Different Ansatz Depths}
    \label{tab:ansatz_depth}
\end{table}

\rev{Next, we investigate the behavior of VarSaw with Global Selective Execution (i.e., sparsity) against VarSaw without Global Selective Execution for a variety of ansatz types and ansatz depths. We perform our simulation based analysis on 6-qubit $CH_{4}$, 6-qubit $H_{2}O$, and 6-qubit $LiH$. The results of our experiments are summarized in Tables \ref{tab:ansatz_type} and \ref{tab:ansatz_depth}. Table \ref{tab:ansatz_type} shows that VarSaw with Selective Execution outperforms VarSaw without Selective Execution for all molecules and ansatz types. For varying depths (Table \ref{tab:ansatz_depth}), we see that Global sparsity helps in all cases but one, where it performs marginally worse compared to the no-sparsity case.}

\begin{figure}
    \centering
    \includegraphics[width = \columnwidth]{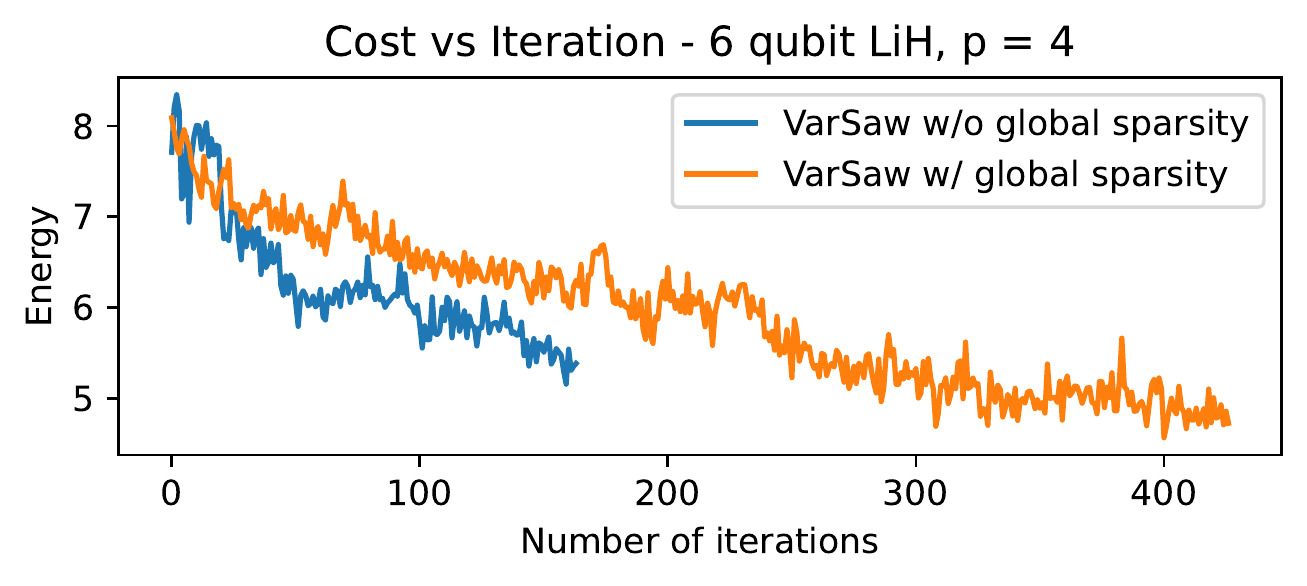}
    \caption{\rev{VarSaw w/ and w/o global sparsity for p = 4}}
    \label{fig:diff_rate_of_conver}
\end{figure}

\rev{We also observe some interesting instances where the rate of convergence of VarSaw with selective execution is poorer than the rate of convergence of VarSaw without selective execution. However, selective execution reduces the per-iteration overhead to such a great extent that the program ends up performing many more meaningful iterations compared to VarSaw with no sparsity, and converges to a lower energy value. A typical plot is shown in Fig. \ref{fig:diff_rate_of_conver}. This is expected since a deeper circuit has a greater number of parameters, and the effect of computing the global distribution using stale parameters is more pronounced - the VarSaw temporal optimization benefit is still seen to have considerable benefit. 
}

\subsection{\rev{Isolated effect of each VarSaw optimization}}

\rev{\emph{VarSaw spatial optimization vs Baseline, for circuit overhead reduction}: Varsaw’s spatial optimization as a circuit reduction scheme does not have any direct impact on the baseline since these subset circuits are not a part of the baseline.}

\rev{\emph{VarSaw spatial optimization vs Jigsaw, for circuit overhead reduction}: Fig.\ref{fig:jig_eval1} showed the impact of Varsaw’s spatial optimization in terms of subset circuit reduction over JigSaw. If we add the constant global component to both (assuming no temporal optimization), then VarSaw is 5x better on average and 12x better for the largest systems.}

\rev{\emph{VarSaw temporal optimization vs Baseline, for circuit reduction}: Fig.\ref{fig:jig_eval2} (secondary axis) shows the benefit of temporal optimization in terms of reduced circuits compared to the baseline. VarSaw only has to execute the Globals around once every 100 iterations. Of course, VarSaw executes the local circuits instead. If we do not use the spatial optimization, then VarSaw would still be much worse than the baseline because the number of local circuits is roughly 6x the baseline circuits. But if we did use temporal optimization as well, then VarSaw is 10x lower circuits than the baseline.}

\rev{\emph{Varsaw temporal optimization vs JigSaw, for circuit reduction}: Similar to above, referring to Fig.\ref{fig:jig_eval2}, VarSaw would execute global circuits only roughly 1\% of the time that JigSaw does. But if spatial optimization is not performed, this benefit is less useful. If spatial optimization is also performed, then VarSaw is 25x lower circuits than JigSaw on average and more than 1000x lower in the largest systems.}

\rev{\emph{VarSaw vs JigSaw, for VQA accuracy}: If there was no runtime or circuit budget, the VarSaw optimizations do not have a direct impact on accuracy over Jigsaw because VarSaw is focused on reducing circuit overhead compared to Jigsaw while employing the same measurement error mitigation. If there was a fixed budget, then VarSaw’s spatial optimization would achieve better accuracy than JigSaw. With both spatial and temporal optimization, this accuracy improvement is 55\% relative to JigSaw’s improvements. If we only had temporal optimization, the benefit would be negligible (as explained earlier, temporal optimization is only beneficial post spatial optimization). The accuracy benefit for spatial optimization alone is around 35\%.}

\rev{\emph{VarSaw vs baseline, for VQA accuracy}: Overall we see a 45\% absolute improvement over the baseline. Without a circuit/runtime budget, this accuracy improvement is independent of the spatial optimization and is unaffected by the temporal optimization, which are primarily circuit overhead reduction techniques.}

\begin{figure}[h]
    \centering
    \includegraphics[width = 0.8\columnwidth, clip, trim={0cm 0cm 0cm 0cm}]{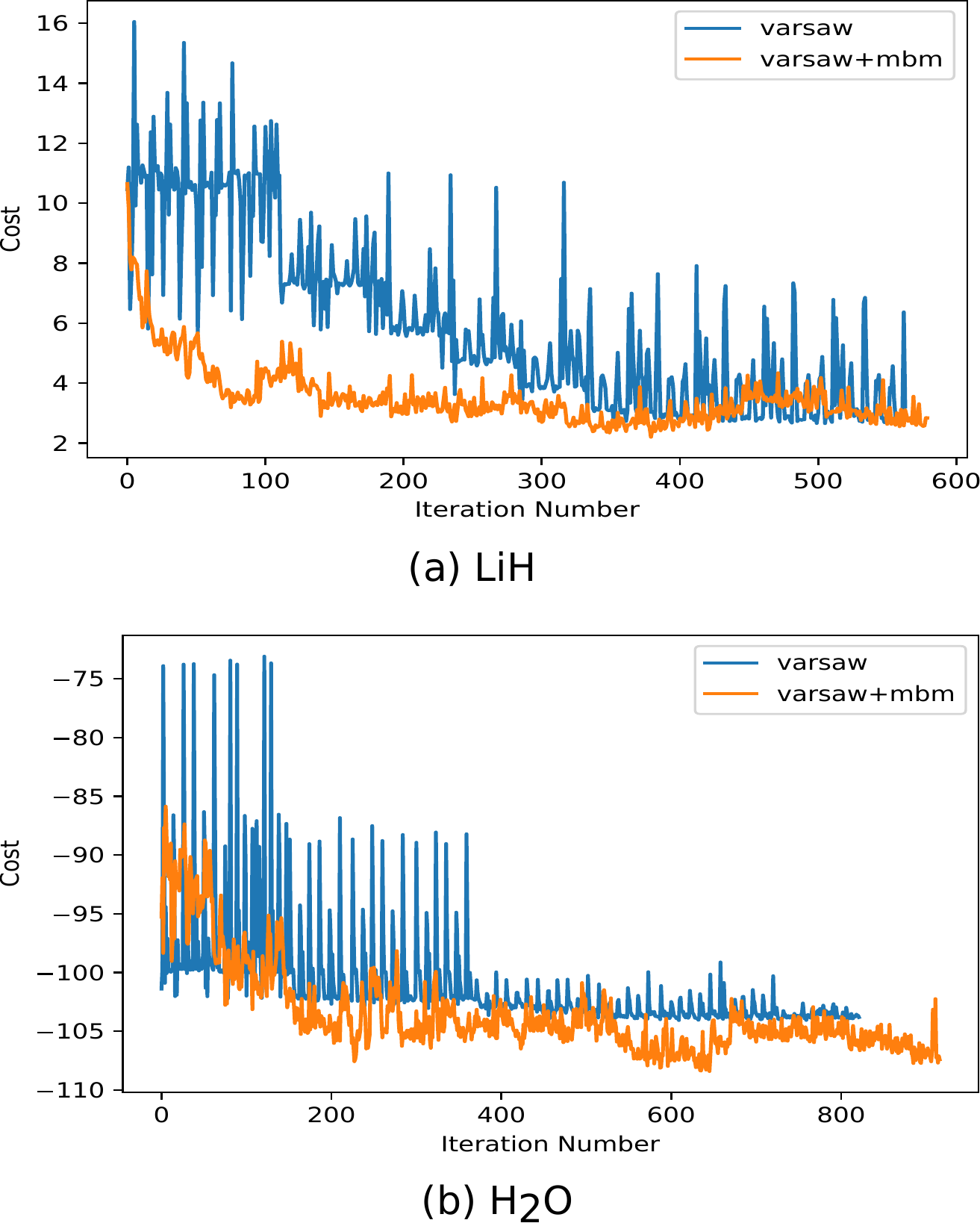}
    \caption{Training VQE with VarSaw and VarSaw+IBM's Measurement Error Mitigation for  $LiH$ and $H_{2}O$}
    \label{fig:varsaw_plus_mbm}
\end{figure}

\subsection{Integration with other mitigation techniques}

{VarSaw can be combined with IBM’s matrix-based complete measurement error mitigation (MBM) \cite{MBM} for higher fidelity than either scheme as standalone \cite{jigsaw}. VQE results are shown in Fig.\ref{fig:varsaw_plus_mbm} for two molecules.
From our evaluations we see that VarSaw+MBM 
improved the VQE estimate by around 10\% for $H_{2}O$ and negligibly, but with less noise, for $LiH$.
A tailored approach to deploying multiple measurement error mitigation techniques is worth pursuing in future work.
} 
\section{Discussion}

\subsection{Why VarSaw works}

The benefit from VarSaw's spatial optimization is fairly straightforward. Eliminating the spatial redundancy in JigSaw subsets is a significant win with no trade-offs. 
The benefit from VarSaw's temporal optimization is more interesting to consider.
Clearly this is a trade-off. 
On the one hand, reducing global executions means that the `weights' used to combine the VarSaw subsets together are stale.
On the other hand, the global executions themselves have inaccuracy due to  measurement errors (among other errors), so fresh global weights does not necessarily imply better overall VQE results.

Therefore, in the presence of non-trivial measurement noise, reducing the impact of the error-prone global runs is advantageous for VQA.
It should also be noted that the VQA ansatz is inherently flexible. 
The rotational gates of the ansatz are tuned by the optimizer to produce a global state that minimizes the overall objective function.
With VarSaw, it is intuitive the optimizer could converge towards the optimal state that minimizes the problem objective along an optimization path which is somewhat less dependent on global weights. This flexibility is offered by VQA and is one of the reasons why there are so many ansatz choices to solve a target VQA problem.
Also important is that the reduced global executions allows VQE to run many more iterations in the VarSaw scenario, so the VQA algorithm is able to search the problem space much faster  even if the weighting mechanism is imperfect.

If measurement noise is negligible, then VarSaw's temporal optimization will not be beneficial. In such a scenario, we cannot expect VarSaw's temporal optimization to outperform the baseline approach. Any additional accuracy in VQA objective estimation that is enabled by the perfect global weights will be lost with VarSaw's stale weights - so this is VarSaw's worst case scenario (which is shown in the top of Fig.\ref{fig:jig_temporal}). However, measurement errors are very significant, so we do not expect this to happen in practice. Additionally, it is possible that device calibration information could be used to influence the use of VarSaw's temporal optimization (or VarSaw as a whole). If some qubits have near-zero measurement errors, then VarSaw, or measurement error mitigation in general, is not required for these qubits.

\subsection{Generalizing VarSaw}

As a quantum application, variational algorithms clearly have the potential for a meaningful shot at a near-term quantum advantage. Thus, we need to enable them towards this goal by building solutions that are as tailored to them as possible.
In this work, we are showing that general-purpose error mitigation techniques, when implemented in a domain-specific manner for VQAs, can provide significant gains beyond the capability of the original technique.
Even more significant is the fact that the original technique (i.e., Jigsaw), when targeting this domain of VQA, is often detrimental at the application level due to its high computational cost (see Fig.\ref{fig:jig_eval2_app} for example), even though it has clear benefits at the circuit level (see Table.\ref{tab:jigsaw_plus_vqe}). We show how its application-level benefits can be transformed  by taking a domain-specific approach to provide substantial gains.
While this work has focused on the intersection of VQA and measurement error mitigation, other optimization techniques, such as circuit synthesis and pulse-level control, can be tailored specifically to VQA.
Similarly, measurement error mitigation can be tailored to other quantum applications like QFT and stabilizers.

\subsection{VarSaw-specific extensions}
\label{FW}

While we have focused on VQE, VarSaw is useful for all VQA problems.
The temporal optimization will produce benefits for all VQAs, but more benefits will be observed for problems that have Pauli terms spread across different measurement bases because this would increase the `Global' cost.
The spatial optimization is beneficial to those VQAs which have Pauli terms spread across different measurement bases (as this would present an opportunity for redundant `Subsets' across the different Pauli terms) but not detrimental otherwise.
These applications include, apart from VQE, time-evolving Hamiltonian simulations that encompass a broad range of algorithms such as the Ising model, Heisenberg model, XY model etc~\cite{2qan}.
Quantitatively evaluating VarSaw on such applications would be useful future work.
There is potential to employ measurement error mitigation only in specific phases of VQA and to only specific terms in the Hamiltonian - i.e., only employ mitigation where it matters most. 
This trade-off of cost vs accuracy is a suitable immediate extension to VarSaw.
Next, in this work, we have focused on VQA with  a hardware-efficient ansatz~\cite{kandala2017hardware}.
A hardware-efficient ansatz can be limited in its capabilities because it is application-agnostic.
In future work, we will look to explore these biases, as well as evaluate the benefits of VarSaw for other ansatz.

\subsection{Related Work}
Prior work VAQEM~\cite{ravi2021vaqem}, dynamically tailors existing error mitigation techniques to the actual, dynamic noisy execution characteristics of VQAs on a target quantum machine. 
While similar in spirit, VAQEM incorporates parameters of different error mitigation techniques into the variational tuning harness - this is specifically suited to mitigation techniques that have tunable parameters and are inherently quantum optimizations (like dynamical decoupling~\cite{DDPokharel_2018}.) 
On the other hand, our work specifically focuses on measurement error mitigation, and optimizes it in ways that is orthogonal to the VAQEM approach.
\rev{Software error mitigation techniques, especially focused on measurement error mitigation have also been proposed in \cite{bravyi2021mitigating, kwon2020hybrid, tannu2019mitigating}. 
Measurement error mitigation techniques tailored to VQAs have been proposed in \cite{barron2020measurement}.
Measurement error can also be mitigated by circuit level advancements.
For example. ~\cite{lienhard2021machine} has proposed a variety of techniques for machine-learning assisted qubit readout.}
Previous work has been done to improve the performance of VQAs using error mitigation techniques. Ref.\cite{kandala2019error} uses Zero Noise Extrapolation (ZNE) to increase the accuracy of VQAs on real quantum hardware. Ref.\cite{wang2021can} shows that error mitigation can improve the trainability of VQAs and overcome the effects of noise-induced barren plateaus.

\section{Conclusion}
In this work, we seek to combat the effect of measurement errors in variational algorithms at reasonable cost.
To do so, we build application-aware optimizations to the general-purpose JigSaw approach for measurement-error mitigation.
VarSaw improves JigSaw~\cite{jigsaw}, specifically in the context of VQAs, by identifying \emph{Spatial Redundancy in JigSaw Subsets} (across VQA Pauli strings) and \emph{Temporal Redundancy in JigSaw Globals} (across VQA iterations) and combating these in novel ways.
In all, VarSaw reduces computational cost over naive JigSaw for VQA substantially while achieving significant improvements in VQA fidelity over both the unmitigated baseline as well as over JigSaw under budget constraints.
Importantly, this work showcases the overwhelming benefits from tailoring state-of-the-art optimizations in a domain-specific manner. Designing JigSaw from the ground up in a VQA-cognizant manner, enables significant benefits. 

\begin{acks}
This work is funded in part by EPiQC, an NSF Expedition in Com- puting, under award CCF-1730449; in part by STAQ under award NSF Phy-1818914; in part by NSF award 2110860; in part by the US Department of Energy Office of Advanced Scientific Computing Research, Accelerated Research for Quantum Computing Program; and in part by the NSF Quantum Leap Challenge Institute for Hy- brid Quantum Architectures and Networks (NSF Award 2016136) and in part based upon work supported by the U.S. Department of Energy, Office of Science, National Quantum Information Science Research Centers. This work was completed in part with resources provided by the University of Chicago’s Research Computing Center. FTC is Chief Scientist for Quantum Software at ColdQuanta and an advisor to Quantum Circuits, Inc.
\end{acks}

\balance
\bibliographystyle{ACM-Reference-Format}
\bibliography{refs}

\appendix
\section{Evaluating different subset sizes} \label{eval5}

{In the JigSaw framework, (i.e., when targeting a single  a single circuit), fixing the subset size naively to 2 may not be optimal. The optimal size would have to be chosen as a trade-off between measurement errors, circuit overheads, reconstruction errors etc. 
In general for JigSaw, smaller subsets means lower measurement error (until some point of saturation) but suffers from higher overheads.}

{On the other hand, with VarSaw,  the smaller subsets are always beneficial---both in terms of accuracy as well as in terms of number of circuits.
There are two aspects here---the first (similar to JigSaw) is that smaller subsets have potentially greater potential for measurement error mitigation (discussed prior). 
The second aspect, and unique to Varsaw, is that smaller subsets actually produce the least number of total circuits for execution. This is because smaller subsets enable considerably larger amounts of commutativity based circuit reduction. 
Thus, Varsaw with 2-qubit subsets is attractive both in terms of maximum error mitigation as well as in terms of lowest number of circuits. Thus, it is the clear choice for our chosen applications. We expect this trend to broadly hold, except for some variation depending on application and machine characteristics.}

\begin{figure}[h]
    \centering
    \includegraphics[width = 0.9\columnwidth, trim={0cm 0cm 0cm 0cm},clip]{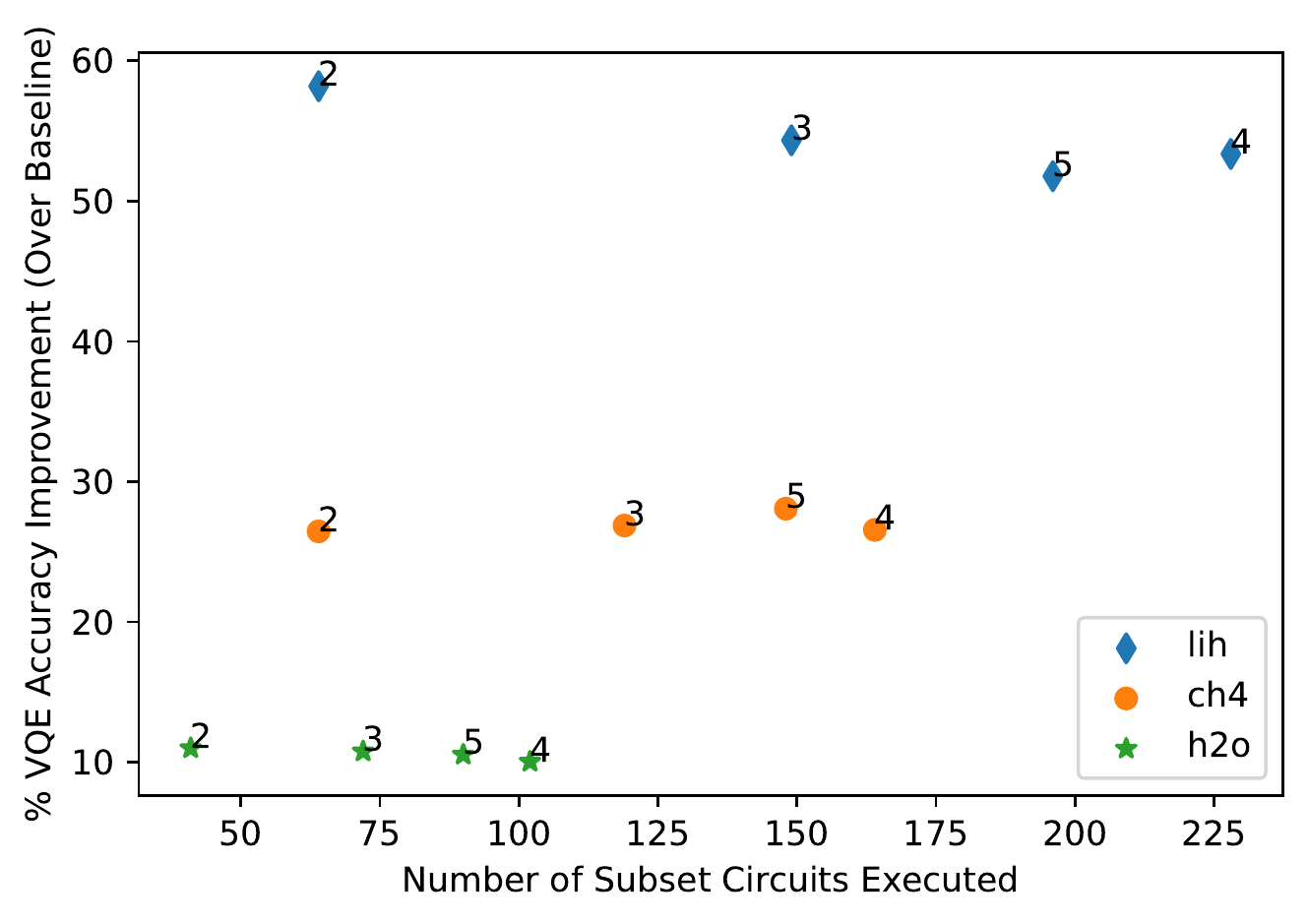}
    \caption{{For one iteration of VQE at optimal parameters, VarSaw is applied on the noisy baseline. While accuracy improvements from subsetting are high, the accuracy variance among small subset sizes is low. But the circuit overhead is substantially lower for smallest circuit subsets.}}
    \label{fig:jig_eval5}
\end{figure}

{In Fig.\ref{fig:jig_eval5}, we run a single instance of the VQE circuit with ideal parameters in noisy simulation and perform VarSaw's measurement error mitigation on top of this.
We plot VarSaw's fidelity improvement over the noisy baseline for 6 qubit LiH, H$_{2}$O, and CH$_{4}$. 
We show results for 4 different subset sizes, all of which produce substantial benefits over the baseline.
For all 3 benchmarks, the accuracy benefits did not vary significantly with subset size - in LiH, the smallest subset performs around 10\% better than other choices. 
In others, the different subset sizes are all attractive in terms of accuracy improvement.
But the important aspect to note is the window size of 2 achieves a considerably smaller number of total circuits  (this number increases with subset size).
Thus, smaller subsets are more attractive for VarSaw.
} 

\section{Performance of Sparsification at Varied Noise}\label{eval6}

{Finally, we evaluate the VQE accuracy benefits from VarSaw's temporal optimization (i.e., adding sparsity)  at different noise levels, to showcase that this optimization is beneficial over a wide noise range.
To do so, we scale the device noise model and use it to run the VQE baseline, VarSaw with No Sparsity and VarSaw with Max Sparsity. Table \ref{tab:sparsity_eval} shows the results for H$_{2}$O-6 molecule. We see that as the noise scales, VarSaw at Max Sparsity always achieves benefits over the baseline and the benefits are always similar (and sometimes better) compared to the  No Sparsity scenario. Thus, not only is the VQE result highly attractive when temporal optimization is employed, it also significantly reduces computational cost. And this trend holds across a variety of noise. Note that when there is no noise at all, then Max Sparsity can perform poorly - this is expected and was discussed in Fig.\ref{fig:jig_temporal} and Section \ref{TRJG}. }

\begin{table}[h]
    \centering
    \begin{tabular}{|c|c|c|c|}
    \hline
         Noise Scale &  Baseline & VarSaw  & VarSaw \\
         & & (No Sparsity) & (Max Sparsity) \\
         \hline
         5 & -89.33 & -91.29 & -91.36 \\
         \hline
         3 & -90.86 & -94.51 & -93.32 \\
         \hline
         1 & -98.28 & -101.73 & -102.73\\
         \hline
         0.8 & -99.45 & -102.95 & -101.08 \\
         \hline
         0.5 & -101.62 & -104.30 & -103.23 \\
         \hline
         0.1 & -104.15 & -104.46 & -104.69 \\
         \hline
         0.05 & -104.57 & -104.75 & -104.77 \\
         \hline
    \end{tabular}
    \caption{{VarSaw with Max Sparsity and No Sparsity compared to the baseline for different noise levels.}}
    \label{tab:sparsity_eval}
\end{table}

\end{document}